\documentclass[12pt]{iopart}
\usepackage{url}
\usepackage{floatflt}
\usepackage{a4wide}                                     
\usepackage[final]{graphicx}
\usepackage{xcolor}
\usepackage{array}
\usepackage{amsbsy}
\usepackage{amssymb}
\usepackage{color}
\usepackage{multicol}
\usepackage{geometry} 
\usepackage{hyperref}
\newcommand{\etmiss}{\mbox{$\rm \Eslash_{T}\!$                        }}
\newcommand{\etmissvec}{\mbox{$\rm \vec{E} \kern-0.6em\slash_{T}\!$                        }}
\newcommand{\Eslash}{\mbox{$\rm E \kern-0.6em\slash$                }}

\begin{document}

\title{The Pursuit of Dark Matter at Colliders - An Overview}
\author{Bj\"orn Penning}

\address{Department of Physics, Brandeis University, 415 South Street Waltham, MA 02453, USA}

\begin{abstract}
Dark matter is one of the main puzzles in fundamental physics and the goal of a diverse, multi-pronged research program. Underground and astrophysical searches look for for dark matter particles in the cosmos, either by interacting directly or by searching for dark matter annihilation. Particle colliders, in contrast, might produce dark matter in the laboratory and are able to probe most basic dark-matter--matter interactions.  They are sensitive to low dark matter masses, provide complementary information at higher masses and are subject to different systematic uncertainties.  Collider searches are therefore an important part of an inter-disciplinary dark matter search strategy. This article highlights the experimental and phenomenological development  in collider dark matter searches of recent years and their connection with the wider field.
\end{abstract}
\submitto{\jpg}
\maketitle

\tableofcontents

\section{Introduction}
We know that about 27\% of our Universe's energy content is composed of an unknown form of matter - dark matter (DM) - that is entirely different from the normal matter which contributes just about 5\%~\cite{Zwicky:1933gu, Ade:2015xua}.  The evidence for DM is strong and consistent across all scales from the galactic to the cosmological~\cite{DMreview, Jungman:1995df}. DM has not yet been directly observed and it will be one of the most important discoveries in fundamental physics. However, all current evidence for DM is purely of gravitational nature. Why then should we use colliders, in particular hadron colliders where particles are mainly produced via the strong force, to search for DM?

The currently preferred `WIMP paradigm' predicts that DM consists of very weakly interacting particles (WIMPs) accessible approximately at the electroweak scale, to account for the observed relic density of particles at the freeze out of the thermal equilibrium in the early Universe~\cite{Srednicki:1988ce, Bertone:2004pz, Feng:2010gw}. 
This electroweak scale energy scale is powerfully probed by the Large Hadron Collider (LHC) at CERN, near Geneva, Switzerland~\cite{Evans:2008zzb}. The freeze-out mechanism requires significant couplings between the dark matter and the standard model (SM), which further motivates searches at a particle collider. Furthermore many `beyond the standard model' (BSM) theories in high energy physics require new particles at the electroweak scale which are either viable DM candidates or might couple to particle DM. The most prominent example of such a theory that connects naturally astrophysical and theoretical motivation is supersymmetry (SUSY). Supersymmetry not only remedies many known problems of the standard model, such as the hierarchy problem, but also provides an excellent DM candidate~\cite{Golfand:1971iw, Clavelli:1970qy, Martin:1997ns}.  Another motivation is the ability to produce and study DM in the laboratory. Collider production implies production of the mediator, i. e. the force carrier that connects the dark sector with the visible sector of the SM. 
This allows to study both particles and their interactions in a controlled environment. The primary experimental signature of DM in a collider detector is missing transverse energy, $\etmiss$, because the particle responsible for DM will escape the detector undetected. The mediator itself might also have additional decay modes into known SM particles. This would lead to modifications in corresponding observables that even at hadron machines can be detected with high accuracy.

\begin{figure}[!htb]
\centering
\includegraphics[scale=.4]{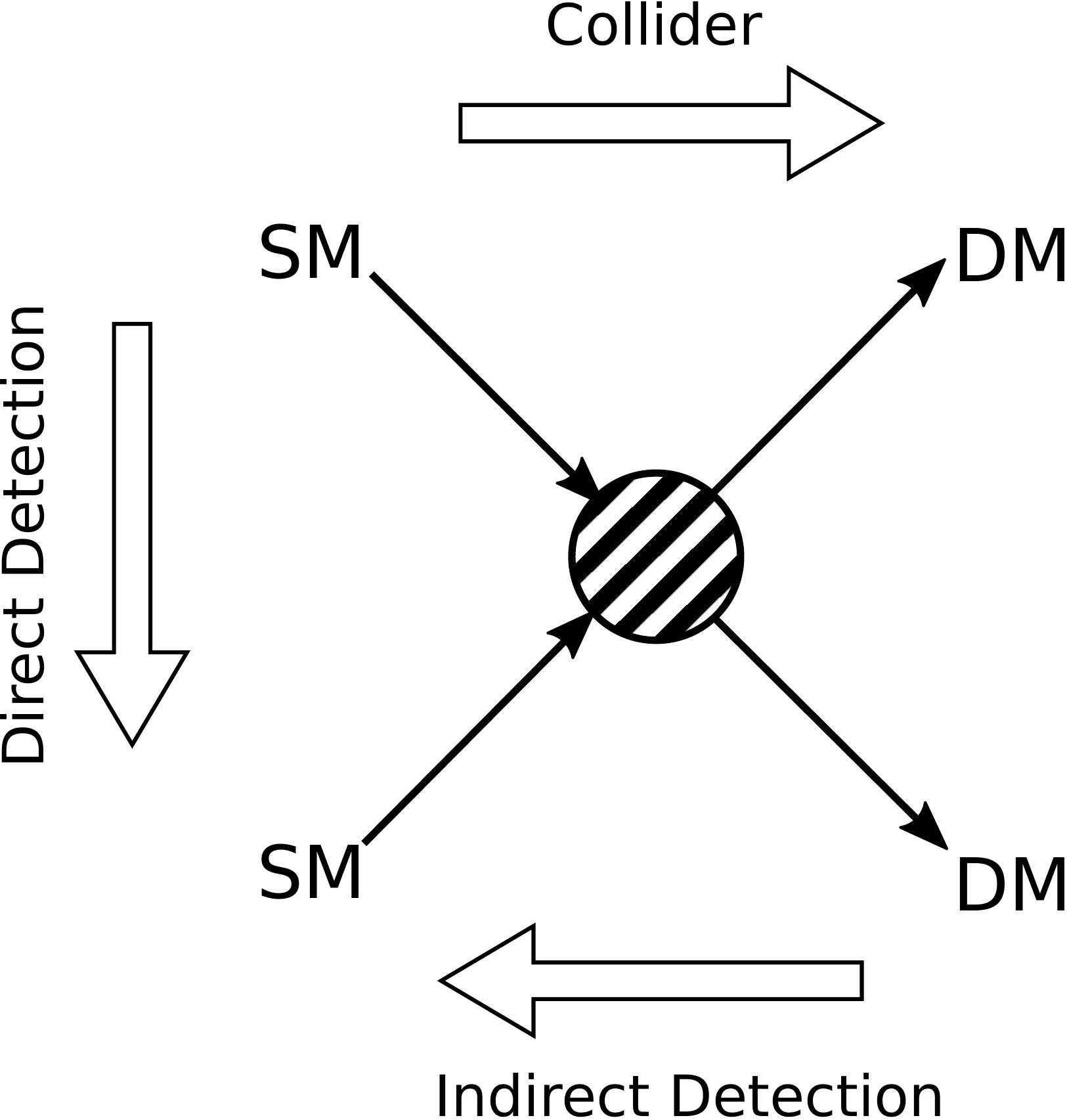}
\caption{Based on the general idea of dark matter interacting with the standard model three detection techniques are possible. These are production at colliders,  scattering from a target material (direct detection) and DM self-annihilation resulting in cosmic rays (indirect detection).}
\label{fig:digraph}
\end{figure}

Finally the complementarity among different approaches to search for dark matter.  
As shown in Fig.~\ref{fig:digraph}, the three different DM detection approaches exploit the production, scattering and annihilation processes provided by a given interaction. Direct and indirect DM searches are affected by large experimental uncertainties in the initial state. These are often of astrophysical nature, such as velocity and density distributions of the DM in the Universe. In contrast the experimental environment at colliders is  well understood. Colliders are also particular sensitive for low DM masses of a few GeV, a region inaccessible to current direct detection experiments and, in contrast to direct and indirect searches, about equal sensitivity to a range of different couplings. However, dark matter and its mediator has to be within the energetic reach of a collider whereas direct and indirect searches can probe dark matter beyond the energies that can be produced currently. Finally for non-WIMP theories please see Ref.~\cite{Feng:2010tg, Kusenko:2013saa}.



\section{Phenomenology}
Collider dark matter phenomenology aims to study viable DM candidates that can be produced at today's colliders and to connect those to different DM searches. A candidate particle needs to be weakly interacting, carry neither  electromagnetic nor colour charge and has to to be stable or with a lifetime of order of the age of the universe or longer. It is desirable that the masses, couplings and production rates reproduce or at least don't conflict with measured DM relic densities and other astrophysical observations. 


\subsection{Effective field theories}
Widely used in early work ~\cite{Beltran:2010ww,Fox:2011pm,Goodman:2010ku} is an effective field theory (EFT) approach. 
The primary assumption of an EFT is that the energy scale of the new physics is large compared to the  energies accessible at the experiment, thus allowing to integrate out the mediator and to parametrise the interaction using effective operators. Table~\ref{tab:operators} lists some of these operators and their coefficients for the most common bilinear covariants of a dirac spinor. 
As long as the momentum transfer in the interaction is sufficiently small compared to the mass of the mediator, the EFT is expected to capture all relevant physics.  Using EFTs, one can constrain the energy scale, $\Lambda$, that is defined by the underlying couplings to the DM, SM particles, and the masses in the model:

\begin{equation*}
\frac{1}{\Lambda^2}=\frac{g_{\rm q} g_{\rm DM}}{m_{\rm med}}
\end{equation*}

Here $g_{\rm q},~g_{\rm DM}$ are the mediator couplings to the DM and SM particles, respectively, and $m_{\rm med}$ is the mass of the hypothetical mediator acting as exchange particle for the force connecting the SM and the DM.

\begin{figure}[h!]
  \center
    \includegraphics[width=0.4\textwidth]{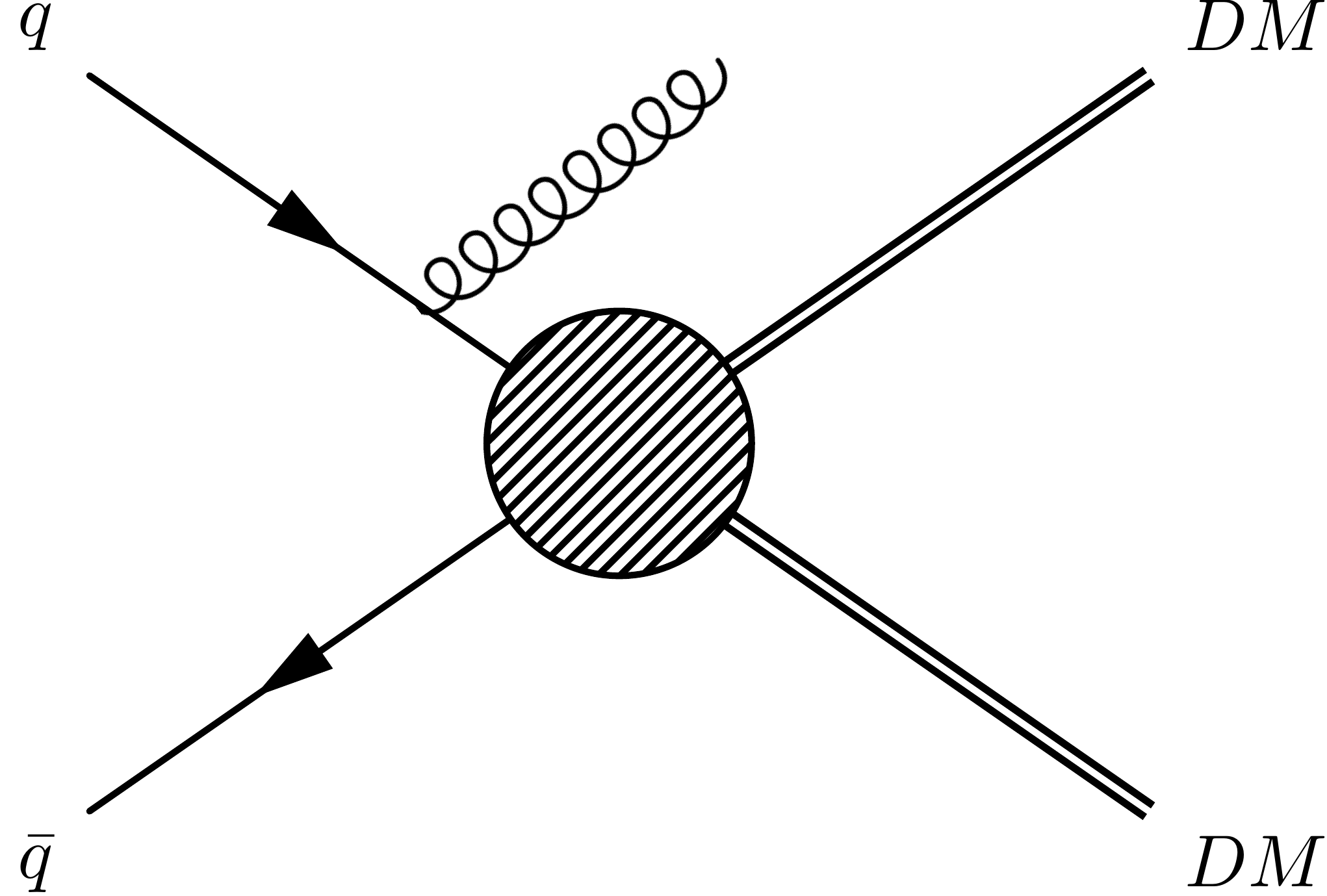}
\caption{Feynman diagrams for the pair production of DM particles for the case of an effective field theory~\cite{Khachatryan:2014rra} \label{fig:feyn_eft}.}
\end{figure}


\begin{table}[h!]
\label{tab:operators}
\begin{tabular}{|c|c|}
\hline
   Name    & Operator   \\
\hline
scalar & $\bar{\chi}\chi\bar{q} q$ \\
pseudo-scalar & $\bar{\chi}\gamma^5\chi\bar{q}\gamma^5 q$ \\
vector & $\bar{\chi}\gamma^{\mu}\chi\bar{q}\gamma_{\mu} q$ \\
axial-vector & $\bar{\chi}\gamma^{\mu}\gamma^5\chi\bar{q}\gamma_{\mu}\gamma^5 q$ \\
tensor & $\bar{\chi}\chi G_{\mu\nu}G^{\mu\nu}$    \\ 
\hline
\end{tabular}
\centering
\caption{Example effective operators between the SM and WIMPs for Dirac fermions. More operators involving complex scalar and real scalar particle DM candidates are described in the literature~\cite{Goodman:2010ku}.}
\end{table}

At a collider, initial states particles can radiate a hard photon or gluon irrespective of the subsequent process as shown in  Fig.~\ref{fig:feyn_eft}.  Besides photons and gluons the high energy at the LHC also allows for the production of heavier particles in the initial-state radiation, for example gauge bosons and the Higgs boson \cite{Carpenter:2012rg, Cotta:2012nj, Crivellin:2015wva,Carpenter:2013xra}.
This initial state radiation (ISR) can be used to record the event because of the presence of the high momentum photon or jet. If a pair of DM particles is produced with sufficient momenta, as required by the EFT approach and many models of new physics, then the ISR object and the DM particles will recoil from each other. Along with the missing transverse energy  $\etmiss$ induced by the non-interacting DM this leads to the distinctive $\etmiss +X$ signature, where $X$ denotes the high momentum SM particle used to trigger the event. The visible and invisible states point in approximately opposite directions of each other. This signature is called `mono-$X$'.  
 
EFT operators can be calculated for a variety of models. 
However, there are several constraints on the validity of EFTs that need to be carefully considered. Because EFTs are non-renormalizable field theories they will break down at larger energies and lead to non-physical results, typical when the energy scales probed are of the order of the mass of the integrated out mediator. Some operators also might break the gauge invariance, particularly for mono-$W$ processes~\cite{Bell:2015sza, Bai:2012xg}. EFTs are as well not able to reproduce the details of many interesting processes such as differences in kinematic distributions or resonant particle production~\cite{Fox:2011pm,Fox:2012ee,Busoni:2013lha, Endo:2014mja}. 
 
Comparisons between collider and direct searches using EFTs are rather straightforward because the latter also use effective theories to extract constraints from their measurements. However, in the case of direct detection the  momentum transfer is  the kinematic energy of the elastic scattering of the DM particles from a nuclei in the non-relativistic regime.  If the mediator mass is of the electroweak scale then for a DM particle of $m_{\rm DM}=100$~GeV the expected energy is at the order $\mathcal{O}({\rm keV})$ and therefore more suitable for EFTs:

\begin{equation*}
E=\frac{1}{2}m_{\rm DM} \times v^2_{\rm DM}  \approx 100 {\rm~GeV} \times 10^{-6} \approx 50 {~\rm keV}
\end{equation*}
The velocity of dark matter is approximated by the velocity of the solar system orbiting the galactic centre, about $220$ km/s. At such low energies an EFT is a very reasonable approximation. Typical momentum transfers at collider on the other hand are $\mathcal{O}(\rm{GeV-TeV})$  and stringent validity constraints need to be applied. 
This undesired behavior is further amplified due to the increasing LHC centre-of-mass energies of the past years.
Finally, the lack of predictive power of EFT kinematics leads to non-optimal searches. EFT interpretations at colliders assume heavy mediators beyond the reach of the collider and their comparison to direct detection DM searches implies sensitivity to arbitrarily small DM masses which is clearly unphysical. 


\subsection{Simplified models}
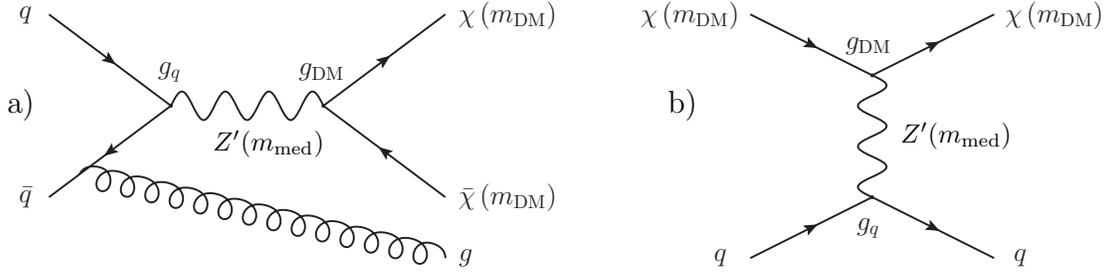
\begin{figure}[ht!]
  \center
    \includegraphics[width=1\textwidth]{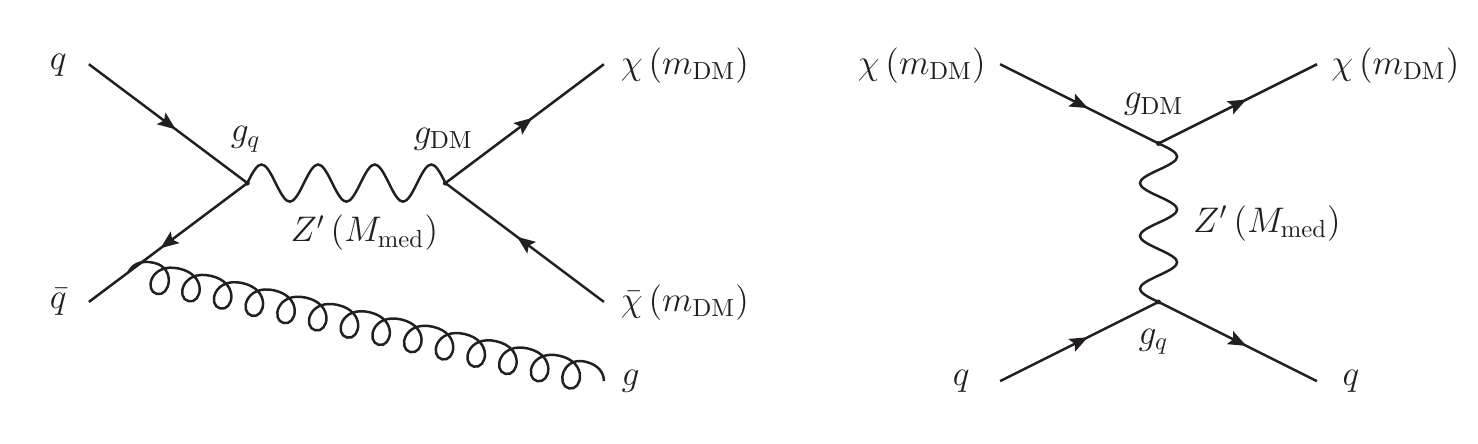}
\put(-420, 73){a)}
\put(-170, 73){b)}
\put(-345, 59){\colorbox{white!30}{\footnotesize $Z'(m_{\rm med})$}}
\put(-85, 62){\colorbox{white!30}{\footnotesize $Z'(m_{\rm med})$}}
\caption{\label{fig:b2_simp}Simplified DM model a) at collider as $s$-channel production process and b) scattering process in direct DM searches.  }
\end{figure}

For a complete description of all processes the effective operator therefore must be replaced by a more complete theory that restores unitarity~\cite{Bell:2016obu, Kahlhoefer:2017dnp, Bell:2015sza, Bai:2012xg}. To do so the mediator particle, the force carrier between the SM and the DM particles, is explicitly taken into account. Only re-normalizable interactions, e.g. dimension 4 or less, are considered~\cite{Buchmueller:2013dya}. These `minimal simplified models'~\cite{Abdallah:2014hon} are designed to reflect the relevant kinematics and final states. This requires the introduction of a number of new parameters: The couplings to the SM ($g_{\rm q}$) and DM sector ($g_{\rm DM}$) and the masses of the mediator ($m_{\rm med}$) and DM ($m_{\rm DM}$). An additional parameter is the width of the mediator $\Gamma_{\rm med}$ which is implicitly taken into account by assuming the minimum decay width of the mediator. This means that the width of the mediator is not a single parameter but is calculated from the other four parameters to allow only decay to the minimal set of particles specified in a given model. This  avoids another potential pitfall of using a single value, that is that the total decay width is smaller than the sum of the partial decay widths into dark matter and quarks. Figure~\ref{fig:b2_simp} shows an example Feynman diagrams of such a simplified model, corresponding to collider production and DM scattering in direct DM searches. The more predictive kinematic and topological distributions of simplified models  allow to develop more sophisticated searches and the application of advanced experimental techniques such as machine learning and particle reconstruction algorithms using jet-substructure.

\subsubsection{$s$-channel models}
The majority of simplified models presently analysed are $s$-channel
production processes as displayed in Fig.~\ref{fig:b2_simp}. The mediator 
between SM and DM particles is usually assumed to be a boson, describing a 
vector or axial-vector interaction in case of spin 1 and 
correspondingly a scalar- or pseudo-scalar interaction if spin 0. 
For spin 1 mediators the couplings to all quarks are assumed to be equal and for scalar mediators proportional to the Yukawa coupling and therefore to the fermion mass~\cite{Abdallah:2015ter} to comply with minimal flavour violation~\cite{Abdallah:2015ter}. Hence these couple most strongly to third generation top and bottom quarks. This mass dependency might help to distinguishing the type of coupling after discovery~\cite{Haisch:2016gry, Haisch:2013fla, Pinna:2017tay, Buckley:2015ctj}.

Typical values of the couplings to quarks chosen are $g_{\rm q}=1.0~(0.25)$ for spin-0 (spin-1) particles, respectively, as recommended by the `LHC Dark Matter Working Group' (DMWG)~\cite{Abdallah:2015ter}. The vertex between DM particle and mediator is always taken to be $g_{\rm DM} = 1.0$. As further detailed in Ref.~\cite{Kahlhoefer:2017dnp} $g_{\rm q}$ should be thought of as a numerical factor such as a mixing angle rather than an independent coupling. The DMWG provides a set of recommendations of models, couplings and parameter space to facilitate comparison between experimental dark matter searches ~\cite{Abdallah:2015ter}.

\subsubsection{$t$-channel models}

In the case that the mediator couples only to one quark and one DM particle, a coloured exchange particle is required that is mediated via a $t$-channel process. These couplings can be universal in terms of colour and generations, but might violate universality with preferred couplings to the first two generations~\cite{Garny:2015wea, Kumar:2013hfa,Agrawal:2011ze, DiFranzo:2013vra, Jacques:2015zha, Bai:2013iqa}. 
Another possibility is that the DM carries the flavour index (flavoured dark matter) ~\cite{Agrawal:2011ze, Agrawal:2014una, Kilic:2015vka, Blanke:2017tnb}, and the strongest couplings might occur to third generation particles (top-flavoured dark matter). Models in this case are reminiscent of `squarks phenomenology' in minimal supersymmetric models and lead to similar collider signatures. A multitude of other very interesting production modes might occur such as pair production of the mediator, spin-2 mediators, scalar DM particles and more. Mediator pair production for example might lead to sensitivities to such signals in dijet searches. To describe them would exceed the scope of this review and we refer the interested reader to the literature, see for example Ref.~\cite{Harris:2011bh, 1988127, Alitti:1993pn, Abazov:2003tj, Aaltonen:2008dn, Aaltonen:2008dn, ATLAS-CONF-2016-029, Sirunyan:2017nvi}. Final state with dilepton resonances are also possible but lead usually to weaker constraints~\cite{Bell:2014tta, Buckley:2015cia, Khachatryan:2016zqb}.


\subsection{Searches for the mediator}

Strictly speaking the models discussed so far imply the discovery of two particles: the dark matter particle and the mediator particle. Indeed colliders are primarily 'mediator discovery' machine often resulting in tighter bounds on the mediator mass than the mass of the potential DM particles.

\subsubsection{Searches using a known mediator}
 A simple but elegant possibility is that the mediator is a known SM particle, often but not exclusively the Higgs boson~\cite{Arcadi:2014lta, Bai:2013iqa, Westhoff:2016rcb, Han:2015dua}. These models are  referred to as `Higgs portal'.  In the case of scalar dark matter, the Higgs-portal interaction can be elementary, compatible with a dark sector that consists of the DM candidate only~\cite{Westhoff:2016rcb, Han:2015dua}.

In the case that the DM mass is equal or less than half of the Higgs boson mass $m_{\rm DM} \leq m_{\rm H}/2$, the Higgs boson decays into pairs of DM particles and corresponding searches are performed~\cite{Khachatryan:2016whc, Aaboud:2017bja, Aad:2015pla}. The Higgs boson's branching fraction into different final states is measured in precision searches that can be combined. By comparing the combination to the SM prediction the rate of decay into DM particles can be constrained. Alternatively analyses are performed that target individual Higgs boson production mode and its decay to invisible particles directly. 

At higher masses $m_{\rm DM}>m_{\rm H}/2$ the Higgs boson becomes off-shell and searches in this region are quite challenging~\cite{Westhoff:2016rcb, Craig:2014lda}. If additional scalar or fermionic mediators exist at the weak scale they could lead to observable signatures, either by mixing with the Higgs boson or by coupling to the massive vector bosons. Possible signatures could be deviations of measured Higgs boson properties from expected SM parameters or multi-lepton signature from pair production of $W$ and $Z$ bosons.

\subsubsection{Searches for an exotic mediator particle}

The primary goal of DM searches is to produce the mediator which then in turn couples to dark matter. The mediator itself might also decay into SM particles, most likely quarks and gluons~\cite{Fairbairn:2016iuf, Fairbairn:2014aqa, Chala:2015ama} that then can be observed as a narrow resonance in the invariant mass of the dijet spectrum. 
Dijet resonance searches have been carried out in many experiments, i.e. UA1, UA2, CDF, D0, CMS and ATLAS~\cite{Harris:2011bh, 1988127, Alitti:1993pn, Abazov:2003tj, Aaltonen:2008dn, Aaltonen:2008dn, ATLAS-CONF-2016-029, Sirunyan:2017nvi}. Final state with dilepton resonances are also possible but lead usually to weaker constraints~\cite{Bell:2014tta, Buckley:2015cia, Khachatryan:2016zqb}. 

Because of the overwhelming $QCD$ multijet background at low energies, increasing trigger thresholds due to additional interactions per event and  larger occupancies for higher centre-of-mass energies and instantaneous luminosities, no single experiment or collider dominates these results. The LHC provides most powerful high mass constraints but allows to constrain lower masses also if the dijet resonance is produced in association with other heavy particles, e.g. the $W$ or $Z$ boson. At low masses results from the Tevatron and UA2 still provide strong limits, but generally the couplings tested are rather large at the order of one~\cite{Dobrescu:2013coa}.  These results are usually interpreted in the light of axial-vector couplings.

\subsubsection{Searches for light mediators at low energy $e^+e^-$ collider, $B$-factories and beam dump experiments}

Very light, sub-GeV scale gauge boson are well motivated by the about $3.5\sigma$ discrepancy between prediction and observation of the anomalous magnetic moment of the muon~\cite{Pospelov:2006sc}. Light mediators might also explain the observed relic densities and small-scale structure problems in cosmology~\cite{Tulin:2013teo, Loeb:2010gj}. 
Because of the large background rates at the LHC, the ATLAS and CMS detectors are generally not sensitive to this mass range with the exception of searches using $B$ mesons. However beam dump experiments and  low energy collider can access this phase space. Additionally rare decays of $\Upsilon$ and $B$ meson searches can be used to set constraints on light scalar (Higgs-like) mediators that is constructed to be in the same multiplet with the DM particle~\cite{Schmidt-Hoberg:2013hba}

Experiments at low energy lepton colliders such as BaBar~\cite{Lees:2017lec} and Belle~\cite{Jaegle:2012sv} provide unique sensitivity to very low DM masses and light mediators of $\lesssim 10$~GeV in a clean experimental environment.
The simplified models  used replace the coupling of the mediator to quarks $g_{\rm q}$  with a corresponding coupling to electrons,  $g_e$. The simplest such model is that the DM couples to the SM via a `dark photon' $A'$~\cite{Holdom:1985ag, Lees:2014xha}. The dark photon is massive mediator from a broken gauge symmetry $U(1)'$ in the dark sector and couples through kinetic mixing with a photon to the electric charge.
In these models $\epsilon$ is used as parameter instead of $g_e$ because of $g_e=\epsilon e q_i$. Scalar and vector type couplings, validity constraints and width are considered as before. The sensitivity for (pseudo-)scalar couplings is small at low energy colliders because of the smallness of the electron Yukawa-couplings~\cite{Essig:2013vha}. While low energy colliders can probe a largely unexplored parameter space only few searches have been performed so far, often with just a fraction of the available data set because of the absence of a dedicated mono-photon trigger. Soon to start high-intensity experiments such as Belle-II~\cite{Abe:2010gxa} will be able to powerfully probe this region.

Fixed-target experiments using beams of protons or electrons can also extended the sensitivity to  light dark matter at sub-GeV scales. In beam dump experiments the DM particles may be produced in collisions with nuclei in the material of the beam dump and then detected downstream of the beam dump. Such experiments have been performed at Fermilab, CERN and SLAC~\cite{Batell:2014mga, Hansl:1978az, Aguilar-Arevalo:2017mqx} with more being planned~\cite{Battaglieri:2014qoa, Alexander:2016aln, Battaglieri:2017aum}. Beam dump experiments also look for dark photons in the context of simplified models. The DM is produced either in the decay of $\pi$ or $\rho$ mesons or the mediator is radiated from the proton and subsequently decays into DM if kinematically allowed~\cite{Aguilar-Arevalo:2017mqx}.





These searches are usually related to the minimal annihilation rate requirement from relic density measurements via the dimensionless variable $y$ that serves as measure of sensitivity for most light mediator searches:

\begin{equation*}
  y=\frac{g_{\rm D}^2 g_{\rm SM}^2 }{4\pi} \left( \frac{m_{\rm DM}}{m_{\rm med}} \right)^2 \geq \langle \sigma v \rangle_{\rm relic} m^2_{\rm DM}
\end{equation*}

Here $g_{\rm D}$ is the dark gauge coupling~\cite{Battaglieri:2014qoa}. 

\subsection{Future developments}
A fairly recent development is the use of two Higgs doublet models (2HDM) ~\cite{Bauer:2017ota, Duerr:2017uap, Wang:2014elb}, a class of simplified DM models for spin-0 mediators that are part of many BSM theories. The dark matter particle is coupled to a new spin 0 mediator and the coupling to the SM happens via the mixing to the Higgs boson or a \mbox{(pseudo-)}scalar partner of the Higgs boson~\cite{Bauer:2017ota}. This allows to satisfy constraints from LHC measurements while still obtaining expected DM densities.
Another field gaining traction is the search for long-lived particles in the context of dark matter. While these searches are not new, the absence of signal in traditional searches led to a re-focus on non-prompt signature and their impact on DM searches~\cite{Strassler:2006ri, Khoze:2017ixx, Baker:2015qna, Bauer:2016gys, Banerjee:2017hmw, Buchmueller:2017uqu}. Additional developments include new production modes~\cite{An:2013xka, Papucci:2014iwa}, dark sectors and non-$\etmiss$ signatures~\cite{Cohen:2017pzm, Cohen:2015toa}, dark-photon searches~\cite{Schmidt-Hoberg:2013hba, Essig:2013lka, Biswas:2016jsh, Wells:2008xg}, novel detectors at the LHC searching for milli-charged or long-lived particles~\cite{Curtin:2017izq, Feng:2017uoz, Haas:2014dda} and beam-dump experiments~\cite{Battaglieri:2014qoa}.

\subsection{Interdisciplinary aspects}

To truly confirm that any discovery of a WIMP candidate at colliders is the particle responsible for dark matter in the Universe would require confirmation from direct or indirect searches. Correspondingly, only an inter-disciplinary combination of several DM detectors is able to measure the properties of dark matter precisely. Therefore it is necessary that the models and methods used at the LHC are transferable to the other DM detection areas, although this is not always trivial~\cite{MarchRussell:2012hi, Profumo:2013hqa}. The most important parameters are predictions of the scattering and annihilation cross sections relevant do direct and indirect detection experiments. If the same model can be utilised to compare different search strategies, then even the absence of a signal in one detection method allows to constrain the permitted parameter space and to optimise searches in the other areas.

Because of the LHC colliding protons, present DM collider searches probe mainly initial states with quarks and gluons. 
Collider searches have different and often smaller systematic uncertainties compared to direct and indirect searches and while the high-mass reach is restricted by the centre-of-mass energy they exhibit very good low mass sensitivities.

\begin{table}[h!]
\centering
\small
\newcolumntype{C}[1]{>{\centering\let\newline\\\arraybackslash\hspace{0pt}}m{#1}}
\begin{tabular}{C{1cm}|C{5.cm}|C{5.cm}|}
    \rotatebox{90}{\it ~~~~EW couplings} \rotatebox{90}{~(Spin 1)~}& {\bf Vector:} \newline $g_{\rm DM} Z'_{\mu} \bar{\chi}\gamma^{\mu}\chi $ \newline Direct detection more sensitive than colliders except at very low dark matter masses.  & {\bf  Axial-Vector:} \newline $g_{\rm DM} Z'_{\mu} \bar{\chi}\gamma^{\mu}\gamma^5\chi $ \newline Direct d`etection and colliders about equally sensitive in different regions of parameter space.
\\ \hline
     \rotatebox{90}{~\it Yukawa couplings~} \rotatebox{90}{~(mass dependent)~} & {\bf Scalar}: \newline $g_{\rm DM} S\bar{\chi}\chi $ \newline Direct detection and collider about equally sensitive in different regions of parameter space. 
 & {\bf Pseudo-Scalar}:\newline  $g_{\rm DM} S\bar{\chi}\gamma^5\chi $ \newline 
 No limits from direct detection, only from indirect. Colliders provide limits comparable to scalar couplings.
\end{tabular}
\caption{Overview of the relative sensitivities of the three main dark matter approaches: direct, indirect and collider searches. To cover the maximal phase space all three approaches are required.\label{tab:interactions}}
\end{table}


Table~\ref{tab:interactions} gives an overview of the individual strengths of the three detection approaches: direct, indirect and collider dark matter searches. While colliders might access DM via all four basic interactions - axial, axial-vector, scalar, and pseudo-scalar - the kinematic reach is limited by the centre-of-mass energy. Direct and indirect searches extend the sensitivity to higher masses of particle dark matter but are not equally sensitive to different couplings. Further details emerge when considering the spin dependence of direct detection searches~\cite{Malik:2014ggr}.


\section{Experimental Environment} 

\subsection{The Large Hadron Collider}

The LHC is the world's highest-energy particle accelerator. The LHC already accomplished  successful data taking periods at centre-of-mass energies of 7 and 8 TeV in 2011, 2012 respectively. The current run, Run 2, started in summer 2015 with an unprecedented centre-of-mass energy of $\sqrt{s}=13$~TeV and nearly three times larger instantaneous luminosity in comparison to the previous runs. The detectors have been upgraded to maintain and improve data taking under these challenging conditions. Rapid accumulation of data is taking place during Run 2 until late 2018. This allows to extend sensitivities to dark matter into yet unexplored regimes with an expected integrated luminosity of about 150 fb$^{-1}$ at $\sqrt{s}=13$~TeV. The entire Run 2 data set will be at least six times larger at almost twice the energy compared to previous data taking periods. Following Run 2, a 18 month shutdown is foreseen to prepare the machine and detectors for even larger luminosities. This will allow the CMS and ATLAS experiments to record a planned $300$ fb$^{-1}$ by the year 2022 in what is called Run 3. Following another two year shutdown the `High Luminosity LHC' (HL-LHC) is the final stage of the LHC. The HL-LHC will deliver up to 3000~fb$^{-1}$ (3 ab$^{-1}$) at potentially $14$~TeV centre-of-mass energies until the late 2030s.

\begin{figure}[h!]
  \center
    \includegraphics[width=0.7\textwidth]{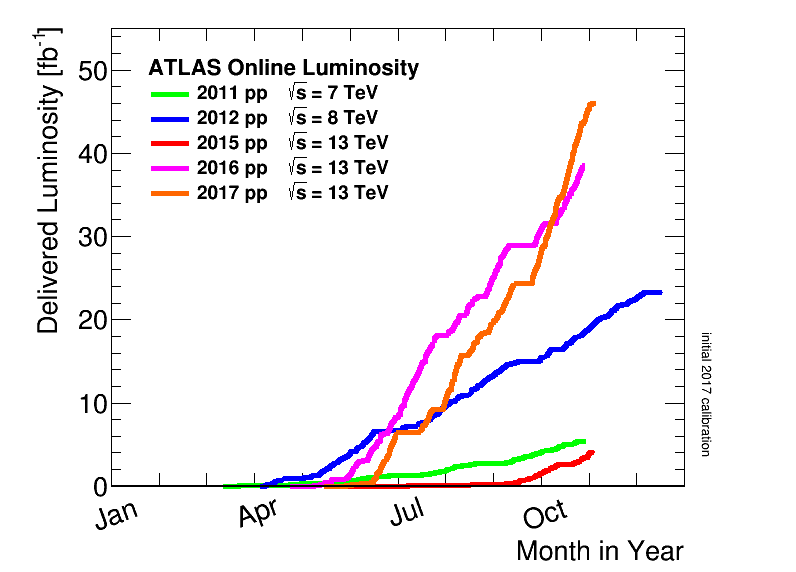}
\caption{\label{fig:lumi} Integrated luminosity of the ATLAS detector between 2011 and late 2017 \cite{atlas_lumi}. The CMS detector data taking is similar.}
\end{figure}

Four detectors are located at the LHC: ATLAS, CMS, LHCb and ALICE. The ATLAS and CMS experiments are multi-purpose experiments and the main focus of this review since the vast majority of DM studies have been performed by these collaborations. The LHCb experiment is designed to study $B$-physics and ALICE to study the collision of heavy ions. Figure~\ref{fig:lumi} shows the integrated luminosity at the ATLAS experiment from 2011 to 2017. 

\subsection{The CMS \& ATLAS detectors}

Within the context of this review we focus on analyses performed at the ATLAS~\cite{Aad:2008zzm} and CMS~\cite{Bayatian:2006zz} experiments. A recent renewed focus on dark photon models will certainly lead to interesting DM related searches by LHCb~\cite{Dettori:2258888} in the near future.

\begin{figure}[h!]
  \center
    \includegraphics[width=0.8\textwidth]{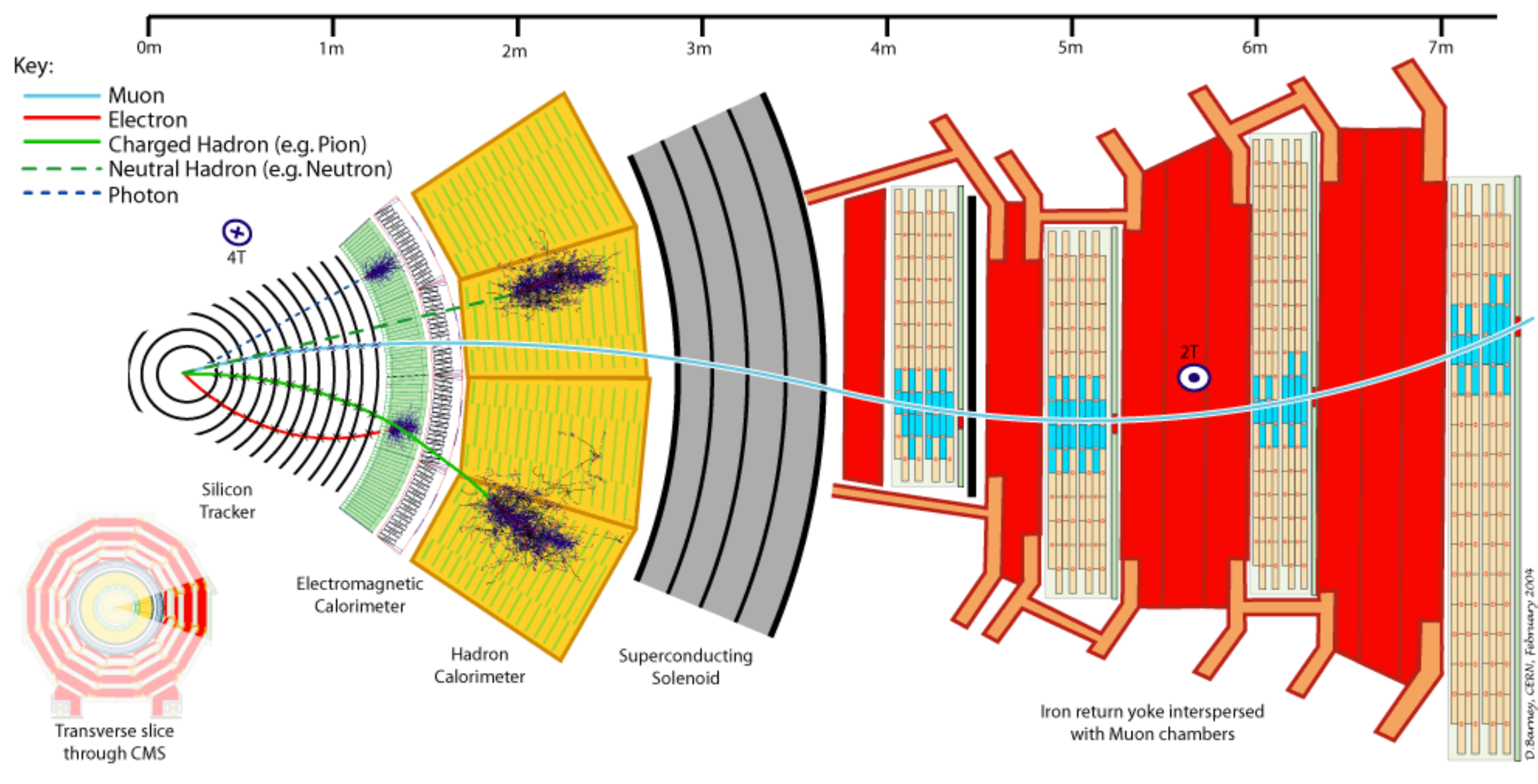}
\caption{\label{fig:b2_cms} Cross section slice of the CMS detector. The detector is segmented into layers of sub-detectors optimised for tracking of charged particles, electromagnetic and hadronic interactions and muon identification. A solenoid creates a 4T magnetic field that aids in the identification of the charge of the particles and momentum measurement~\cite{Barney:2120661}.}
\label{fig:detector}
\end{figure}

The LHC provides proton collision at four different experiments at different interaction regions. The  layout of ATLAS and CMS is typical for all-purpose detectors, consisting of tracking detectors to measure the position of charged particles passing through, calorimeters measuring the energy of electromagnetic and hadronically interacting particles by absorption and finally a muon spectrometer. 
Magnetic fields are generated by a solenoid magnet in CMS and a solenoid and a toroidal magnet in ATLAS. The $\vec{B}$-field enables precise momentum measurements of charged particles. Figure~\ref{fig:detector} shows the cross section of the CMS detector. A sophisticated trigger and read out system classifies particles of interests such as high momentum or rare decays in real time to reduce the data volume suitable for storage and electronic processing.   
 

\subsection{Missing transverse momentum}

\begin{figure}[h!]
  \center
    \includegraphics[width=0.8\textwidth]{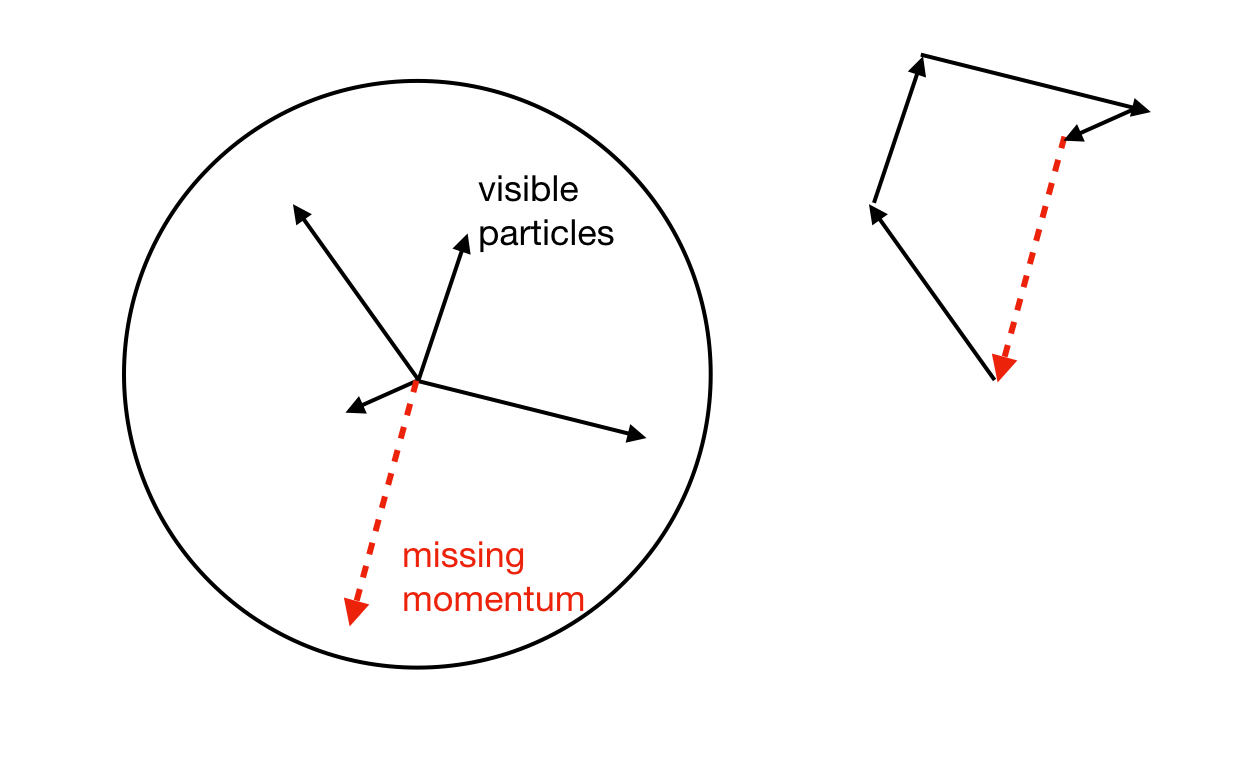}
\caption{Transverse plane of the detector where the black arrows represent visible particles, the red arrow represents an invisible particle. If the vector sum of the visible particles is not vanishing then the corresponding momentum can be associated with the `missing transverse momentum'\label{fig:met}.}
\end{figure}

The particles colliding and therefore their decay products have no significant momentum in the direction transverse to the beam and the vectorial sum of the transverse components must vanish. `Invisible' particles such as neutrinos and WIMPs escape undetected and thus cause an imbalance in the transverse plane. This imbalance is associated with `missing transverse energy', $\etmiss$,  or more correctly `missing transverse momentum' that provides the main signature for most collider DM searches. The missing transverse energy is defined as the negative vector sum of the transverse energy of all of the detected particles. 

\begin{equation*}
\etmissvec=-\sum_i \vec{E}_{\rm T}^i
\end{equation*}
The principle of \etmiss reconstruction is sketched in Fig.~\ref{fig:met}.

\section{Collider Searches for Dark Matter}\label{sec:collider_dm}

Collider dark matter searches can be broadly distinguished into two categories: Searches in final states with and without dark matter itself. 
In the first case, if the mass of the dark matter is small compared to mediator, then the WIMP pair is boosted opposite to the direction of the visible particle(s), leading to the `mono-$X$' signature, $\Delta \phi(\etmiss, X)\approx \pi$. Here $X$ can be a multitude of particles such as $\gamma, g, q, W, Z, H$ and others. Today's searches extend to larger particle multiplicities and hence `mono-$X$' has become technically a misnomer but is still widely used to refer to collider dark matter searches. For DM heavier than the mediator, the mediator becomes off-shell, resulting in weaker constraints from colliders due to kinematically suppressed production cross sections~\cite{Buchmueller:2014yoa}.

The latter case are searches for the mediator. An example is the dijet analysis looking for a narrow peak in the invariant mass of the two jets, $m_{jj}$. 

So far no significant excess has been observed, placing significant constrain on a large number of dark matter models and masses.  More exotic DM candidates such as unparticles~\cite{McDonald:2008uh, Aliev:2017bme}, microscopic black holes~\cite{ArkaniHamed:1998nn, Dimopoulos:2001hw}, Kaluza-Klein~\cite{Cheng:2002ej} states and others have been discussed in the literature and corresponding searches were performed~\cite{Sirunyan:2017onm, Sirunyan:2017anm, Aad:2014gka} but will not be discussed in detail

\subsection{Mono-X searches}

\begin{figure}[h!]
  \center
    \includegraphics[width=0.3\textwidth]{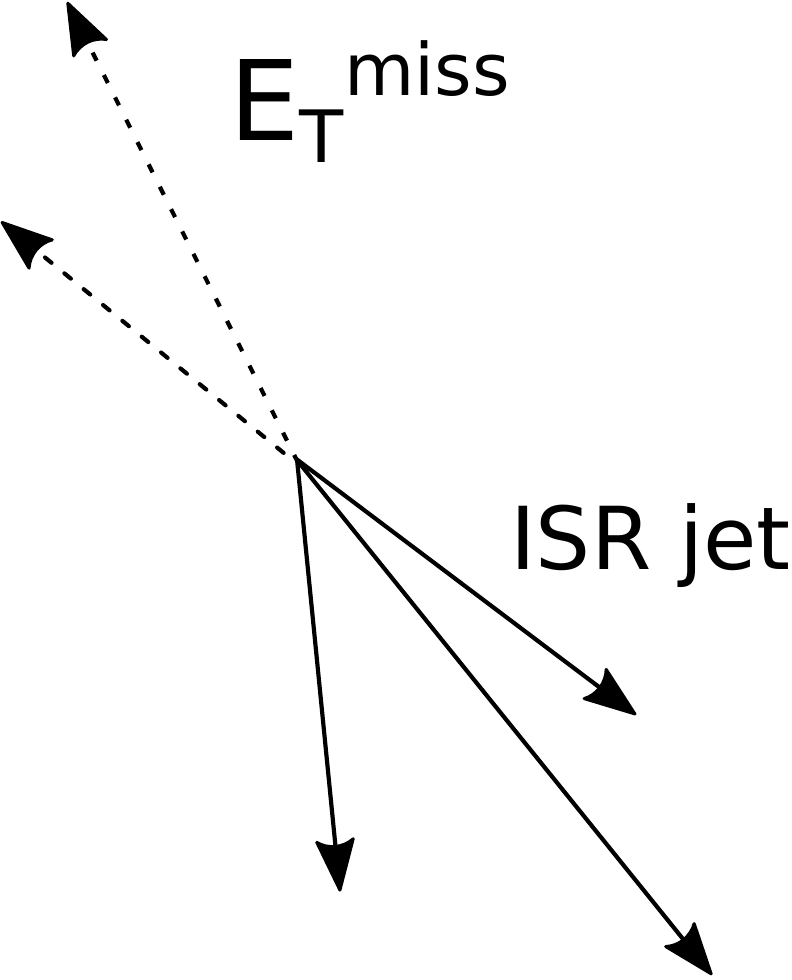}
\caption{\label{fig:dm_signature} The mono-$X$ signature. The generated WIMP pair is boosted into the same direction but opposite to the visible hadronic particles.}
\end{figure}

This `mono-X' signature, shown in Fig.~\ref{fig:dm_signature}, has been pioneered using a single photon in the final state by the experiments at the `Large Electron Positron' (LEP) collider~\cite{Fox:2011fx, Abdallah:2003np} in searches for large extra dimensions~\cite{Randall:1999ee}, which itself is not a DM search. %
The first search using photons at a hadron collider was performed at the Tevatron $p\bar{p}$ collider by the CDF collaboration at $\sqrt{s}=1.96$~TeV \cite{Acosta:2002eq}. D{\O} performed the very first mono-jet analysis also in the search for large extra  dimension~\cite{Abazov:2003gp}. By the early 2010s using LHC data at a centre-of-mass energy of $\sqrt{s}=7$~TeV the potential of this signature for DM searches has been fully recognised and different final states analysed, extending experimental signatures to larger jet-multiplicities and more complex final states involving heavy quarks, gauge, and Higgs bosons.

These early dark matter searches at the Tevatron and LHC were true mono-X searches in the sense that those employed final states with exactly one reconstructed high momentum jet or photon that is required to be approximately back to back to the \etmiss. Dominant background sources are $Z(\to \nu \nu)$ production and $W(\to \ell \nu)+{\rm jets}$ production where the lepton fails reconstruction. These searches were updated with increasing centre of mass energies and datasets, the energy increase of the LHC from $7$ to $8$ TeV led to about one order of magnitude improvement in sensitivity on the traditional WIMP-nucleon--cross-section. The development of simplified models that provide more predictive kinematics~\cite{Buchmueller:2014yoa}, the danger to misinterpret EFTs perturbative bounds and the fact that even in EFTs the majority of events contain more than one high-$p_{T}$ jet~\cite{Haisch:2013ata} lead to the inclusion of higher jet multiplicities and new final states towards the end of the 8 TeV run.

\subsubsection{Searches for dark matter with photons (mono-$\gamma$)}
As indicated by the name mono-photon searches look for a single high-momentum photon produced in association with the DM.  They were the  first dedicated DM searches at particle colliders and continue to play a leading role. Because of the relative strength of the electromagnetic coupling constant $\alpha_{\rm EM}$ compared to the strong coupling constant $\alpha_{\rm S}$, the sensitivity of mono-$\gamma$ analyses is reduced in comparisons to hadronic searches. However, they constitute a particularly clean channel to study DM interactions.

\begin{figure}[!htb]
\centering
\includegraphics[scale=.4]{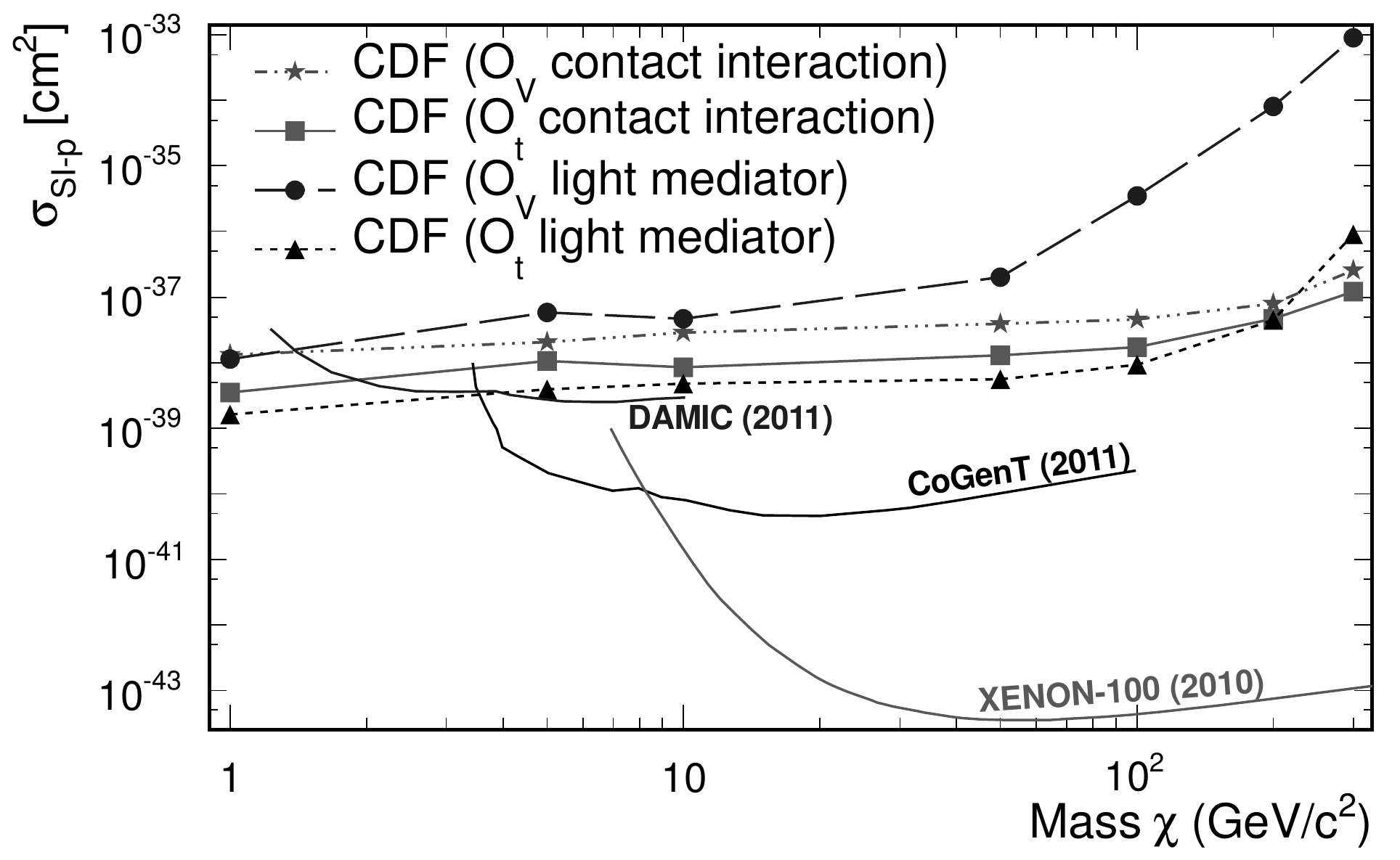}
\includegraphics[scale=.34]{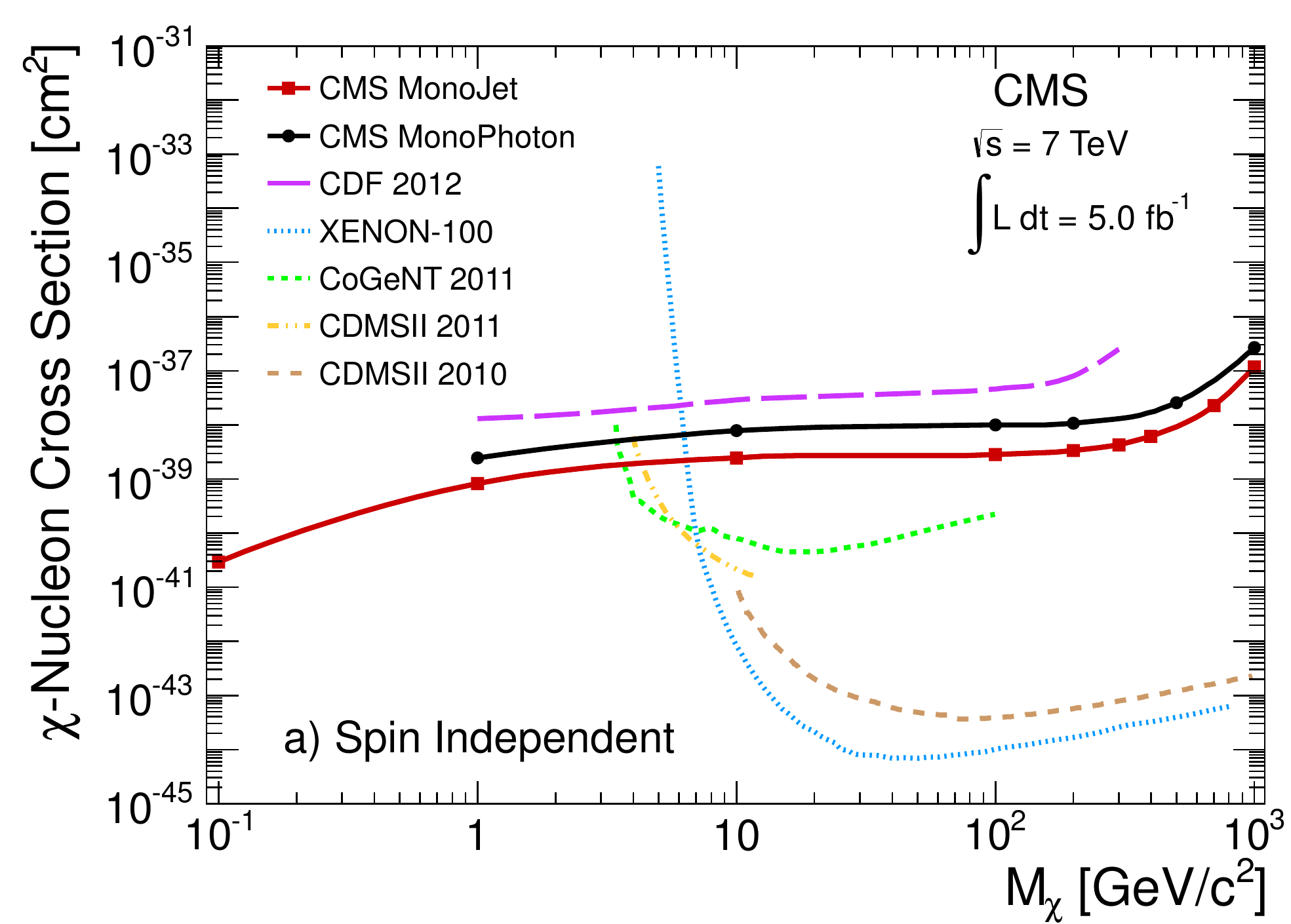}
\caption{First Tevatron mono-photon (left) and LHC mono-photon/jet  (right) dark matter constraints ~\cite{Chatrchyan:2012tea, Aaltonen:2012jb}  compared to results from DAMIC \cite{Barreto:2011zu}, CoGeNT \cite{Aalseth:2010vx}, XENON-100 \cite{2010PhRvL.105m1302A}, SIMPLE \cite{Felizardo:2011uw} , CDMS \cite{Ahmed:2010wy} and COUPP \cite{Felizardo:2011uw}. These searches use effective theories.}
\label{fig:early_dm}
\end{figure}

The analyses performed by ATLAS and CMS select events based on the momentum of the photon and $\etmiss$. Inclusive and exclusive signal and control regions with increasing tight requirements on $\etmiss$ are used to estimate and constrain background processes and to enrich the signal. The backgrounds are dominated by $Z(\to \nu\bar{\nu})+{\rm jets}$ and $W+\rm{jets}$ production.  Figure~\ref{fig:early_dm} shows the first search at the Tevatron at $\sqrt{s}=1.96$~TeV in $p\bar{p}$ collision, soon followed by  LHC searches at $\sqrt{s}=7$~TeV in $pp$ collisions~\cite{Chatrchyan:2012tea, Chatrchyan:2012me, Aaltonen:2012jb}.  Most recent searches analyses the full 2016  LHC dataset of $13$~TeV data~\cite{Sirunyan:2017ewk, Aaboud:2017phn} and are presented in Fig.~\ref{fig:boost} and Sec.~\ref{sec:results}.

\subsubsection{Searches for dark matter with jets (mono-j/V)}
\begin{figure}[tbp]
  \centering 
  \includegraphics[width=.7\textwidth]{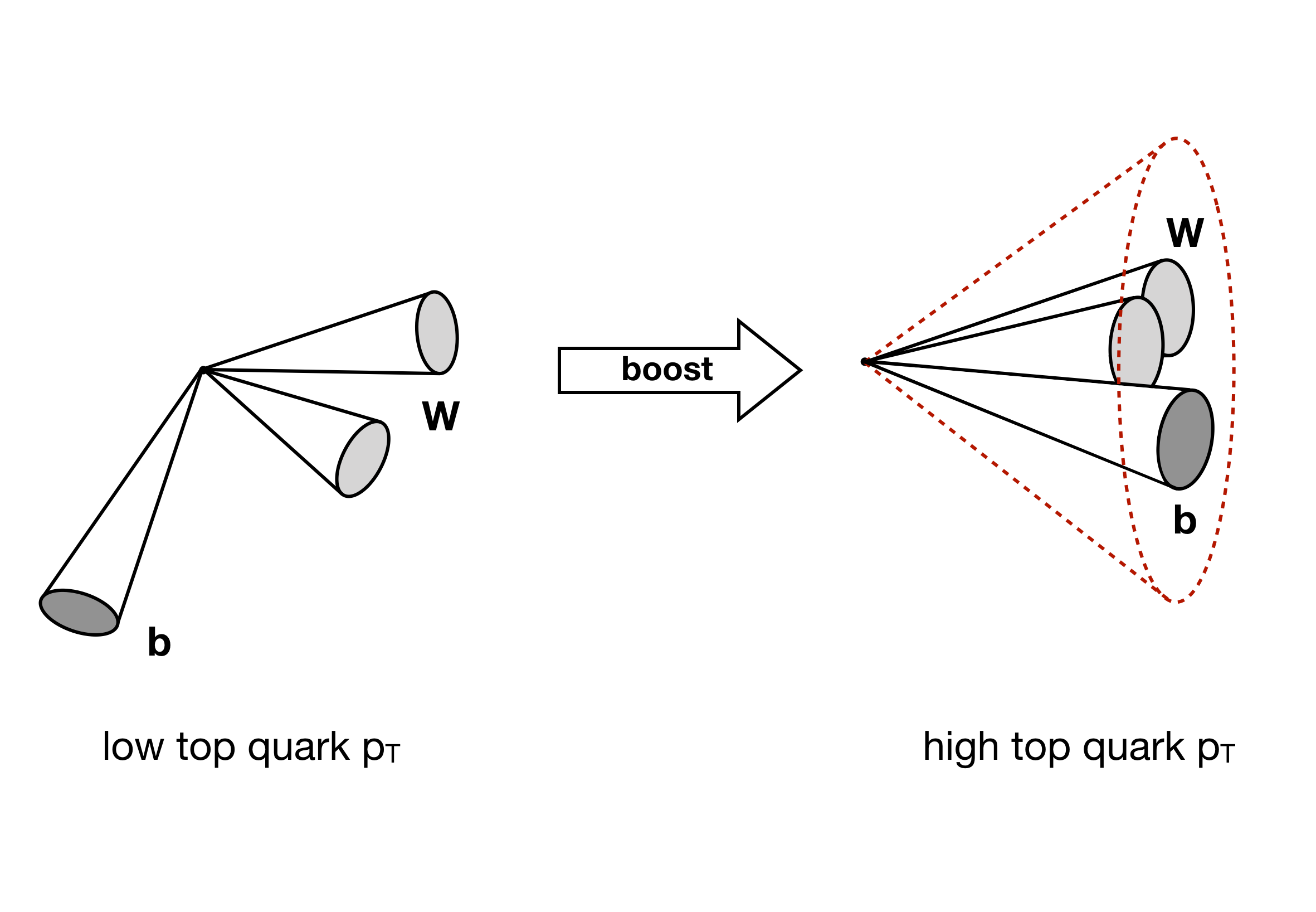}
  \caption{
   Schematic representation of the decay of a boosted top quark. The decay products are detected with momentum in the same direction as the momentum of the initial particle.}
  \label{fig:boost}
\end{figure}

Searches in hadronic final states constitute the majority of collider DM searches. Mono-jet (mono-$j$) searches select events based on the presence of at least one high momentum jet and $\etmiss$ and no reconstructed leptons. This includes searches with jets from the hadronic decay of $W$ and $Z$ bosons (mono-$V$). At the energies provided by the LHC their decay products can be  boosted, collimating the daughter particles in a large radius jet with a substructure reflecting the origin process.~\cite{Shelton:2013an, Carpenter:2013xra} as illustrated in Fig.~\ref{fig:boost}.

\begin{figure}[!htb]
\centering
\includegraphics[scale=.136]{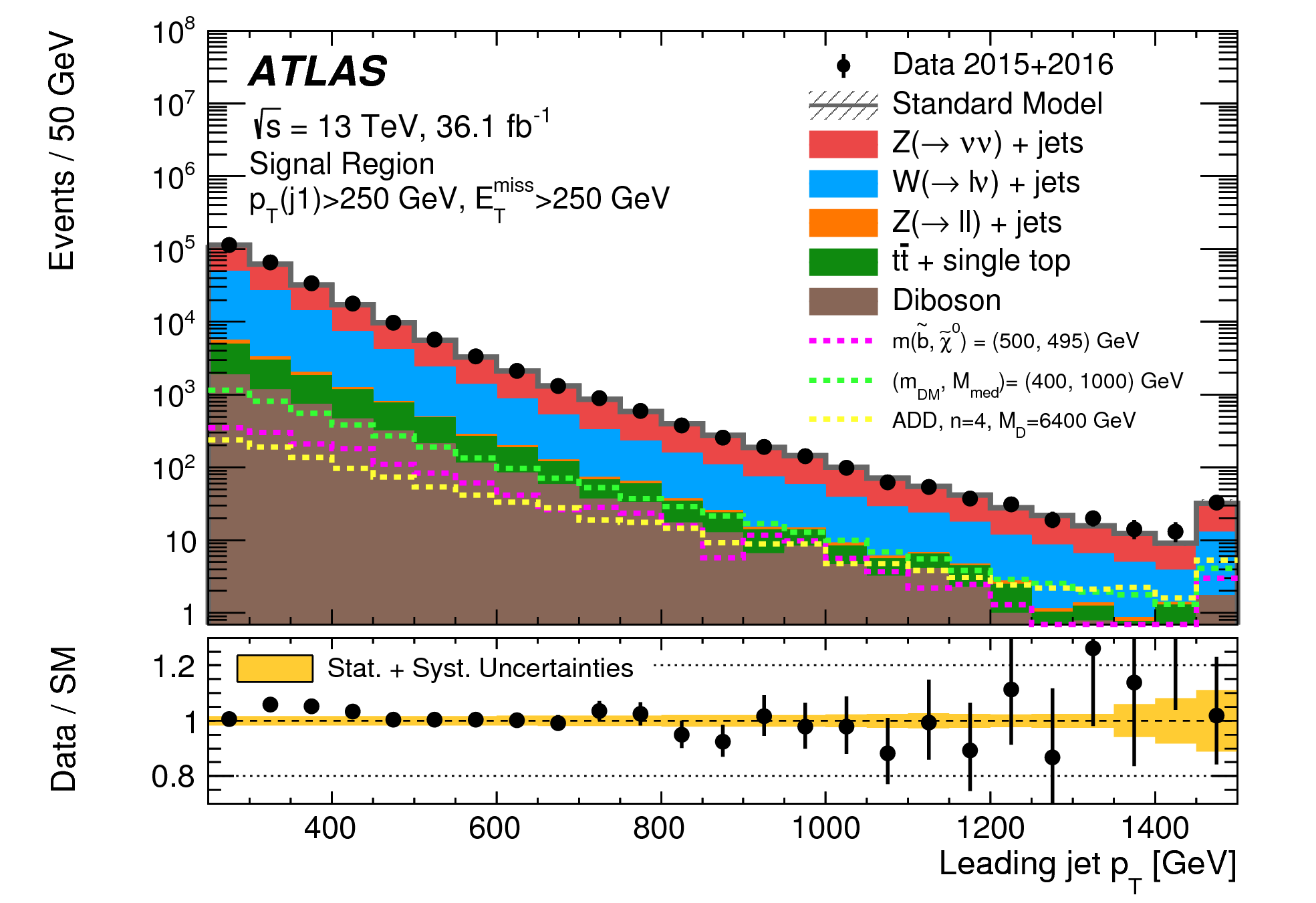}\hfill
\raisebox{-.026\height}{
\includegraphics[scale=.062]{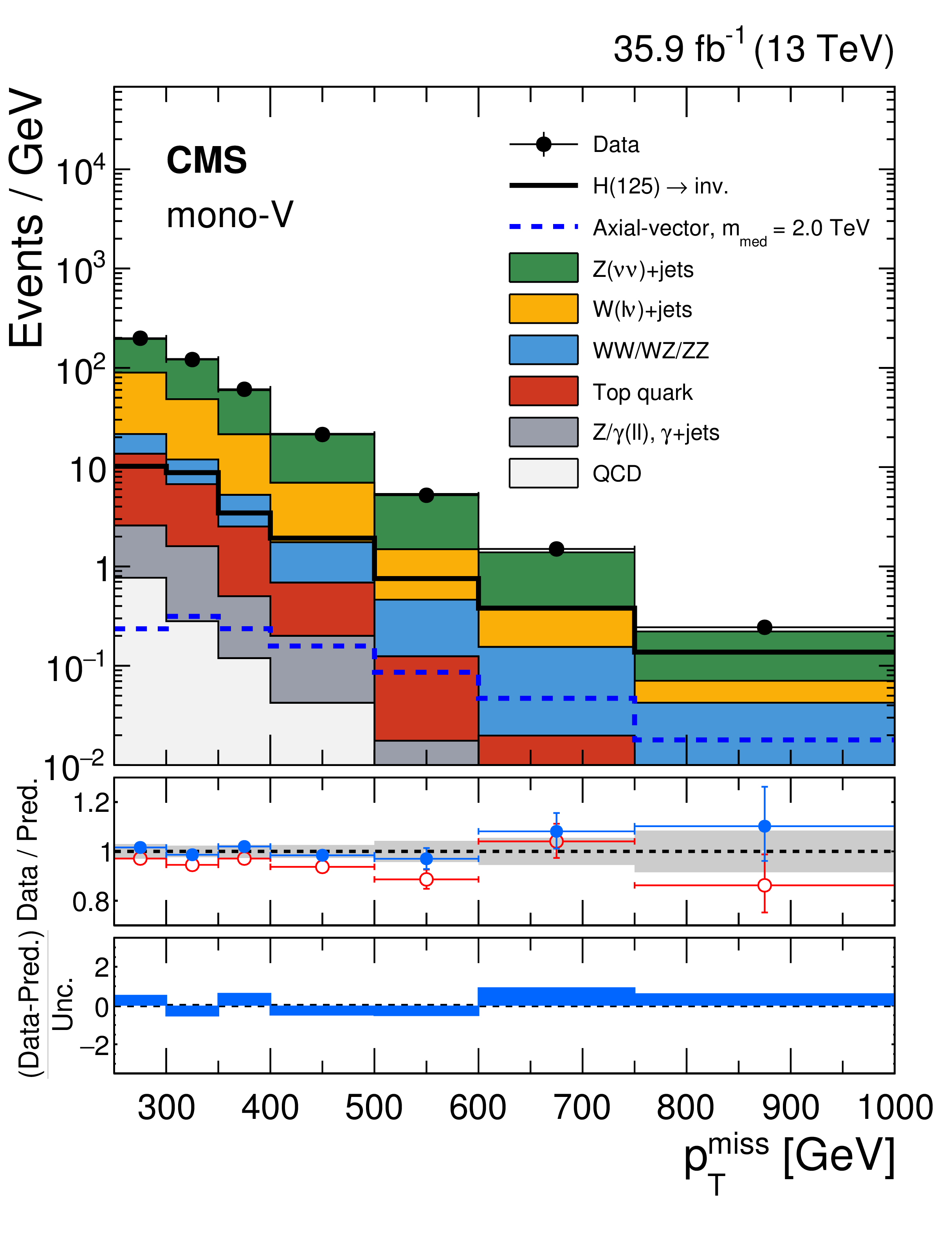}}
\caption{Observed missing momentum distributions in the ATLAS mono-jet signal region ~\cite{Aaboud:2017phn} (left). CMS additionally studies the mono-$V$ process~\cite{Sirunyan:2017jix} (right). Expected background contributions are given by the solid histograms and the markers gives the observed distribution in $36$~fb$^{-1}$ of data.}
\label{fig:monojetV}
\end{figure}

For low DM masses the dominant signal production mode is mono-jet whereas for higher masses mono-$V$ gains importance. Backgrounds are dominated by  $Z(\to \nu\bar{\nu})+{\rm jets}$ and $W+\rm{jets}$ processes and are estimated separately for mono-$j$ and mono-$V$ channels in a number of dedicated data control regions. Limits on mono-$j$ and \mbox{mono-$V$} production are derived based on the $\etmiss$ distribution that lead to the currently strongest constraints on DM from collider searches. Corresponding $\etmiss$ distributions from ATLAS and CMS are presented in Fig.~\ref{fig:monojetV}.

\begin{figure}[!htb]
\centering
\includegraphics[scale=.41]{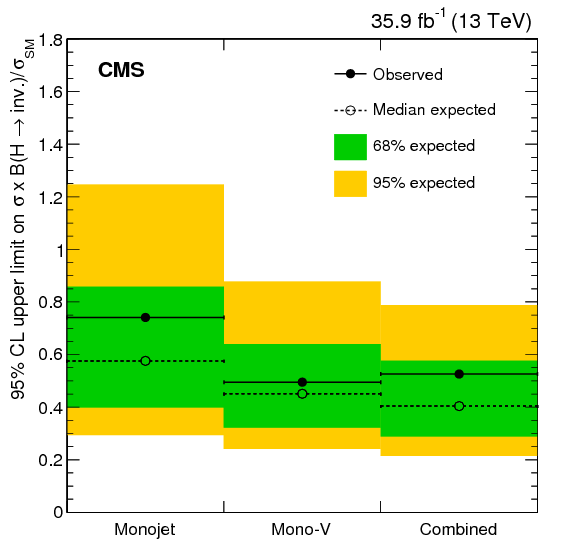}
\caption{Expected (dotted black line) and observed (solid black line) 95\% CL upper limits on the invisible branching fraction of the 125 GeV SM Higgs boson. Limits are shown for the mono-jet and mono-$V$ categories separately, and for their combination~\cite{Sirunyan:2017jix}.}
\label{fig:monojetHinv}
\end{figure}

Another possible DM discovery channel is the decay of the Higgs boson to DM particles. The result of the mono-$j/V$ search is then interpreted in terms of upper limits on the decay of the Higgs boson to invisible particles, $H \to {\rm inv}$. \mbox{Mono-$j/V$} searches constrain the $H\to {\rm inv}$ branching fraction to about $50\%$ of the total decay width of the Higgs boson, see Fig.~\ref{fig:monojetHinv}.  While dedicated measurements set more stringent constraints of about $25\%$,  mono-jet analyses contribute to the combination of $H \to {\rm inv}$ because they are based on an independent data sample.
Throughout the $13$~TeV LHC run the mono-$j/V$ search provides the strongest constraints on simplified dark matter models.~\cite{Sirunyan:2017hci, Aaboud:2017phn}

\subsubsection{Searches for dark matter with heavy quarks}

\begin{figure}[h!]
  \includegraphics[width=.9\textwidth]{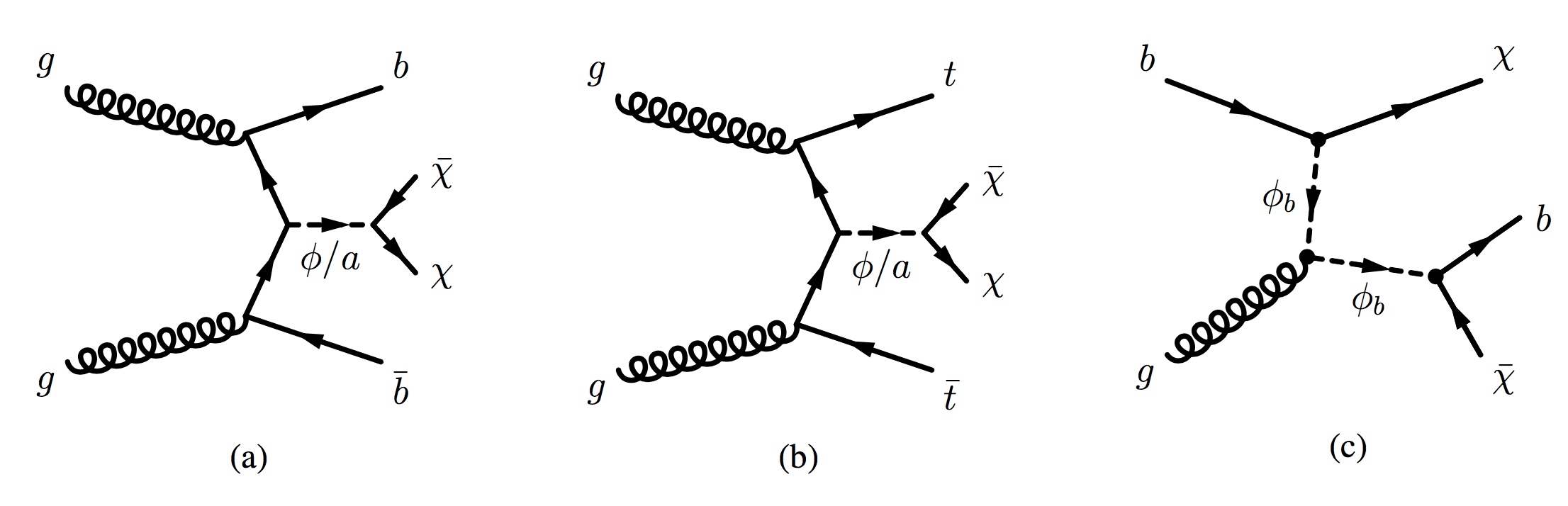}
  \caption{Leading order Feynman diagrams for spin-0 mediator associated production with top and bottom quarks. (a) colour-neutral spin-0 mediator associated production with bottom quarks; (b) colour-neutral spin-0 mediator associate spin-0 mediator associated production with top quarks; (c) colour-charged scalar mediator model decaying into a bottom quark and a DM particle \mbox{$b$-FDM}~\cite{Aaboud:2017rzf}. }
  \label{fig:feynman_heavy}
\end{figure}

If DM  is produced by the exchange of a spin-$0$ scalar or pseudo-scalar colourless mediator, or a colour-charged  scalar mediator~\cite{Agrawal:2014una, Artoni:2013zba}, then minimum flavour violation assumption implies Yukawa-like couplings. Yukawa couplings are proportional to the fermion mass, causing preferred coupling to heavy quarks.  Leading order Feynman diagrams are shown in Fig.~\ref{fig:feynman_heavy}.

DM plus heavy quark searches were first analysed using the LHC 8 TeV dataset Ref.~\cite{dmhf, Khachatryan:2015nua}. These searches already probed DM+$b\bar{b}$ and DM+$t\bar{t}$ final states, used simplified DM model and employed discriminating variables besides missing transverse momentum. The range of production modes in combination with the possible top-quark decays leads to variety of possible final states. Additional $\etmiss$ is produced in the decay of the $b$-quarks, low $\etmiss$ final states can be addressed using the dileptonic decay of the $t\bar{t}$ quark system and $b$-tagging helps to discriminate $b$-jets from the overwhelming $QCD$ multijet background. 

\begin{figure}[h!]
\centering
\includegraphics[scale=.5]{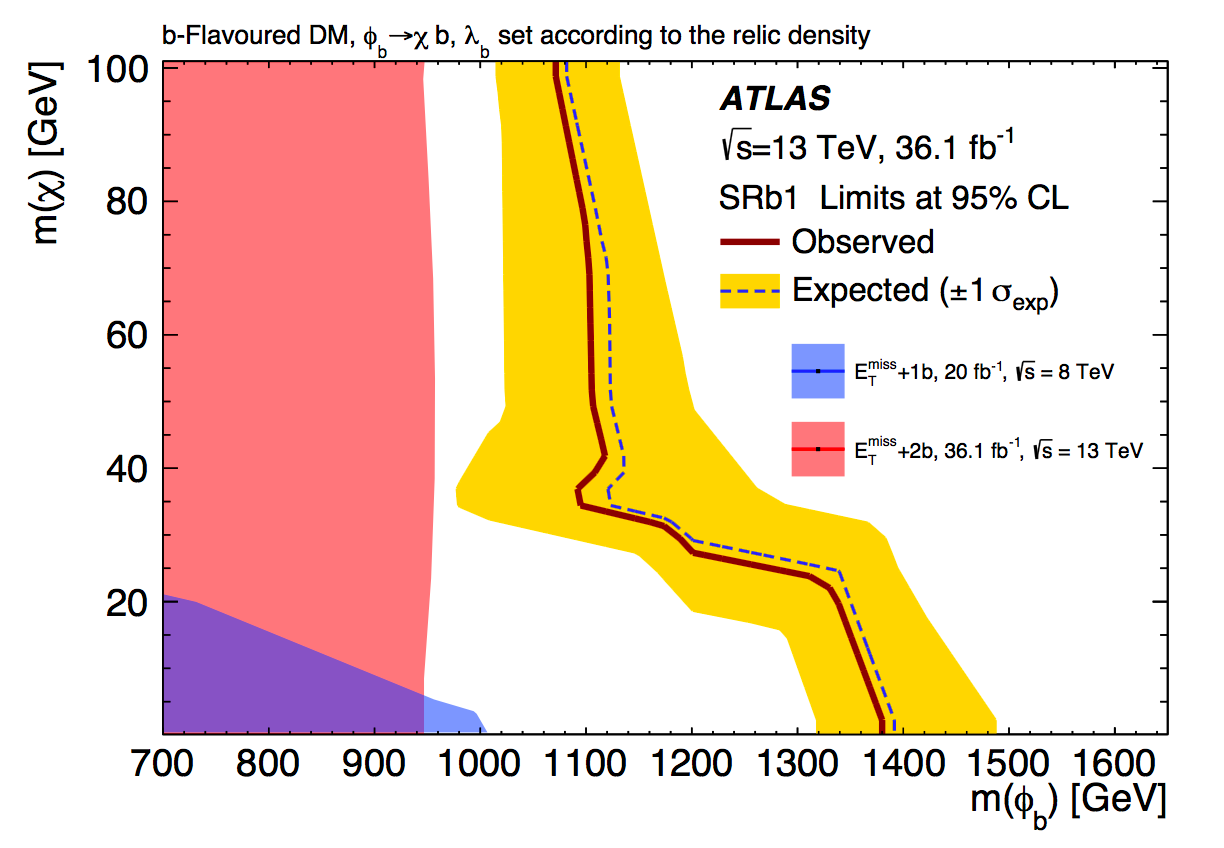} 
\caption{ Exclusion limits for colour-charged scalar mediators (b-FDM) as a function of the mediator and DM masses for 36.1 fb$^{-1}$ of data. The limits are calculated at 95\% CL. The solid (dashed) line show the observed (expected) exclusion contour for a coupling assumption yielding the measured relic density. No uncertainties on the LO cross-sections are considered for this model. The results are compared with the ATLAS search for $b$-FDM models~\cite{Agrawal:2014una}, represented by the blue contour, and the ATLAS search for direct sbottom pair production~\cite{Aaboud:2017wqg}, represented by the red contour~\cite{Aaboud:2017rzf}.}
\label{fig:bfdm}
\end{figure}

Data are usually collected using $\etmiss$  and di-lepton triggers. Dominant backgrounds are $Z(\to \nu\bar{\nu})+{\rm jets}$ and $W+\rm{jets}$ production for the low $\etmiss$ and $t\bar{t}$ and $Z(\to \nu\bar{\nu})+{\rm jets}$ for the high $\etmiss$ signal regions. The background processes are estimated using signal depleted control regions in data and MC simulations. Because the missing transverse momentum provides less discrimination power in heavy quark searches,  distributions such as the opening angle between the $b$-quarks, or the angle between \etmiss and $b$-quarks are used~\cite{Aaboud:2017rzf}. The CMS collaboration also uses a kinematic fit to the top quark and W bosons masses to reconstruct jet momenta, energy, and resolutions. This `top tagger' improves purity of the selected events and sensitivity for DM plus top quark models~\cite{Sirunyan:2017xgm}. Searches for colourless spin-0 particles are just reaching sensitivity based on DMWG recommendations~\cite{Abdallah:2015ter}. Figure~\ref{fig:bfdm} shows an example in which the limits derived from collider measurements are compared to astrophysical indirect measurements.

For a DM particle of approximately $35$ GeV, as suggested by the interpretation of data recorded by the Fermi-LAT Collaboration~\cite{Daylan:2014rsa, Fermi-LAT:2017yoi}, mediator masses below 1.1 TeV are excluded at 95\% CL by the $b$-FDM, a model containing a colour-charged scalar mediator model decaying into a bottom quark and a DM particle~\cite{Agrawal:2014una}.



\subsubsection{Searches for DM with Higgs bosons}

\begin{figure}[h!]
\centering
\includegraphics[scale=.25]{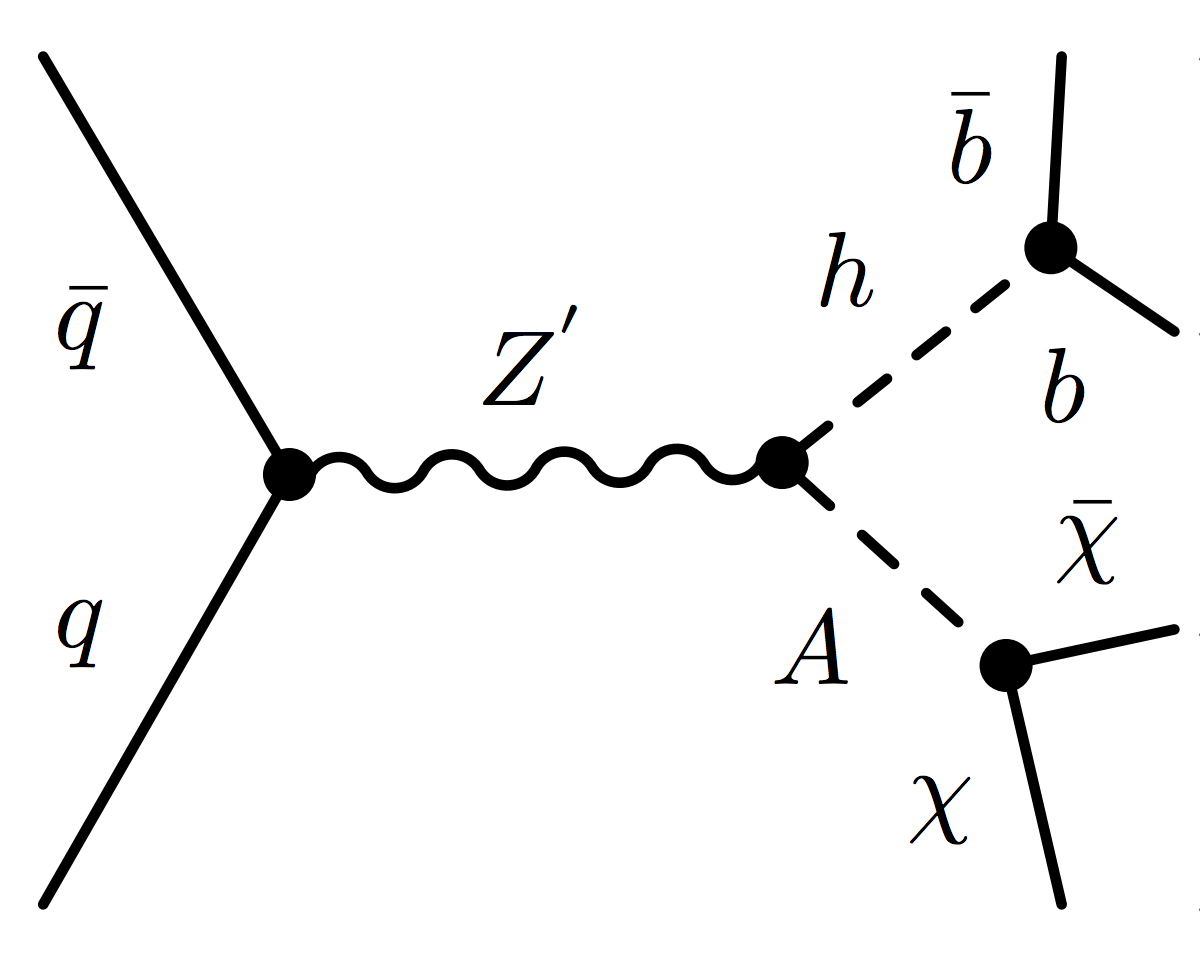} 
\caption{Feynman diagram for the production of DM $\chi$ in association with a SM Higgs boson arising from a $Z'-2HDM$ model~\cite{Duerr:2017uap}. }
\label{fig:monoH}
\end{figure}

The mono-Higgs channel combines  characteristics of the mono-photon and \mbox{mono-$V$} searches. Initially proposed in Refs.~\cite{Carpenter:2013xra, Berlin:2014cfa, Petrov:2013nia} they are implemented in EFTs and simplified models. Most simplified mono-H models either invoke a $Z'$ type vector mediator in which the $Z'$ radiates the SM-like Higgs boson or a $Z'-2HDM$ model where the dark matter couples to a heavy pseudo-scalar particle in an extended Higgs boson sector. The leading order Feynman diagram is shown in Fig.~\ref{fig:monoH}.

Leading searches are performed in the $H\to b\bar{b}$ decay with a branching fraction of about $\sim 57\%$ and in the $H\to\gamma\gamma $ channel that occurs to about $\sim 0.23\%$. The former is the Higgs boson's dominant branching fraction with large backgrounds, the latter is one of the more rare yet cleaner decays. Other analysis channels are mono-$H(\to WW)$, mono-$H(\to ZZ)$ and mono-$H( \to \tau \tau)$.

Analyses using the $b\bar{b}$ final state apply tight selection criteria to reduce the  $QCD$ multijet background and employ $\etmiss$ or the invariant mass of the di-$b$-jet system as the discriminating variable. Backgrounds are dominated by $W/Z+{\rm jets}$ production. If the Higgs boson is produced with large transverse momentum then the decay products, similar to the `mono-$V$' case, are merged into a single large radius jet that contains two $b$-quark jets. This boosted behavior can be exploited using `Higgs-tagging' algorithms~\cite{ATL-PHYS-PUB-2015-035}.

Searches in the diphoton channel  employ looser selection criteria and are often derived from the corresponding SM $H \to \gamma \gamma$ measurement. In these measurements the backgrounds are parametrised as fit from data~\cite{Aaboud:2017uak}. Current searches in the $\gamma \gamma$ \cite{Aaboud:2017uak,Sirunyan:2017hnk} and $b\bar{b}$~\cite{Sirunyan:2017hnk,Aad:2015dva,Aad:2015yga} channels do not exhibit any statistically significant fluctuation.

Recent developments in models with an axial-vector mediator~\cite{Kahlhoefer:2017dnp} lead to an increased interest in simplified models with more than one mediator~\cite{Ghorbani:2015baa, Choudhury:2015lha, Duerr:2016tmh, Ghorbani:2016edw}. Although it is necessary to introduce one additional parameter, these models are attractive due to their rich phenomenology. For example, if the spin-1 mediator decays visibly, one may obtain a mono-$Z'$ signature~\cite{Autran:2015mfa, Bai:2015nfa} while visible decays of the spin-0 mediator may lead to a signal from a `dark Higgs boson`~\cite{Duerr:2017uap}.


\subsection{Dijet searches}

The dijet final states offers a largely model-independent sensitivity to new physics by searching for a narrow resonance on the exponentially falling $QCD$ multijet background. Dijet searches hence have been performed at most hadron colliders~\cite{Harris:2011bh, 1988127, Alitti:1993pn, Abazov:2003tj, Aaltonen:2008dn, Aaltonen:2008dn, ATLAS-CONF-2016-029, Sirunyan:2017nvi}.

A new $s$-channel produced particle that decays into a pair of jets can be observed as a narrow resonance in the invariant dijet mass at about the particle's mass ~\cite{Dobrescu:2013coa}. Depending on the new particle's branching fractions, the dijet channel might be the most viable and straightforward way to discover new physics. Interpreting the new particles as the spin-1 mediator of simplified models, the dijet searches place tight constrains on the mediator mass in DM searches~\cite{Sirunyan:2016iap, Aaboud:2017yvp, An:2012va}. In order to still obtain significant mono-$X$ production, the coupling of the mediator to quarks has to be considerably smaller than the coupling to dark matter. Because of the large QCD background at low masses, high $p_{\rm T}$ jets are needed to maintain manageable trigger rates in proton collision. Hence these searches are typically sensitive to mediator masses of $\mathcal{O}(100-1000)$~GeV. The low mass reach can be extended by taking into account the angular distribution of the jets or requiring associated production with heavy gauge boson~\cite{Aaboud:2017yvp, Sirunyan:2017ygf}

\begin{figure}[h!]
  \centering 
  \includegraphics[width=.6\textwidth]{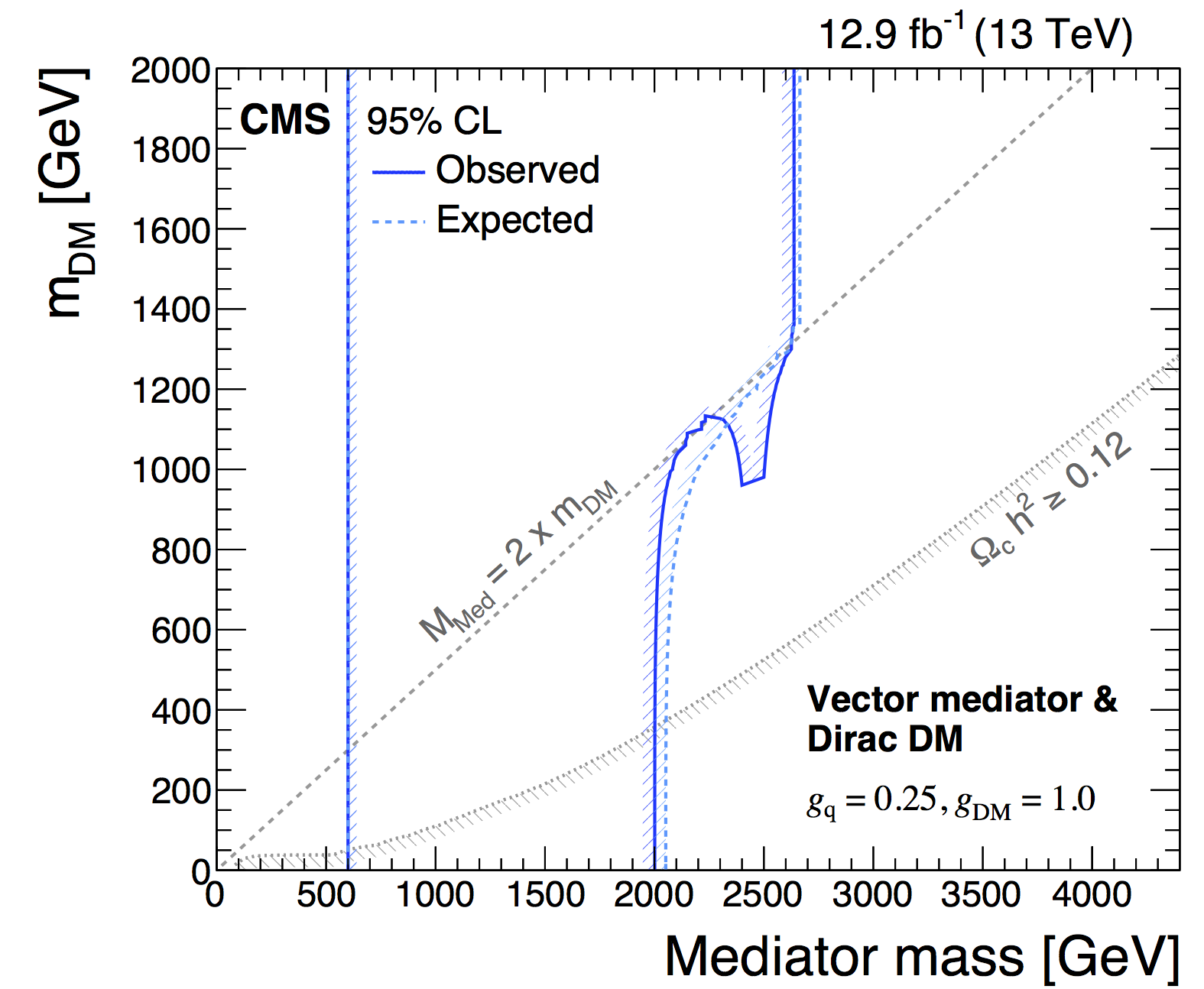}
\caption{ The 95\% CL observed (solid) and expected (dashed) excluded regions in the plane of dark matter mass vs. mediator mass, for a vector mediator,  compared to constraints from the cosmological relic density of DM (light gray)
determined from astrophysical measurements~\cite{Spergel:2006hy, Ade:2013zuv}. Figure from Ref.~\cite{Sirunyan:2016iap}.}
  \label{fig:dijet}
\end{figure}

At typical coupling choices of $g_{\rm q}=0.25$ and $g_{\rm DM}=1.0$ dijets searches place the strongest constraints between about 500 GeV up to almost 3 TeV in mediator mass for almost all value of DM masses as seen in Fig.~\ref{fig:dijet}. If the couplings to quarks are smaller, even for larger couplings to dark matter, these constraints can be relaxed.



\subsection{Supersymmetric searches}
Supersymmetry (SUSY) ~\cite{Golfand:1971iw, Clavelli:1970qy, Neveu:1971rx, Volkov:1973ix, Wess:1973kz} is an extension of the standard model that assigns to each SM particle a superpartner with  a spin differing by half a unit. If SUSY is realised in nature it could solve open question in particle physics such as the hierarchy problem and grand unification, and it provides a viable DM candidate. In most $R$-parity conserving SUSY theories the lightest stable supersymmetric particle (LSP) is the DM candidate~\cite{Farrar:1978xj}. In many minimal supersymmetric models (MSSM) the LSP is the lightest neutralino ($\tilde{\chi}^0$), a physical superposition of the superpartners to the weak hypercharge, the $W$s and the Higgs (Bino, neutral Wino and Higgsinos)~\cite{Munoz:2017ezd}. There are other viable SUSY candidates, for example the right-handed sneutrino~\cite{Hagelin:1984wv}. A comprehensive review of the SUSY model space and analyses is not possible within the scope of this review and we refer to the literature~\cite{Munoz:2017ezd, Jungman:1995df, Olive:2016efh, Belanger:2015epa, Bechtle:2015nta}.

\begin{figure}[h!]
  \centering 
  \includegraphics[width=.98\textwidth]{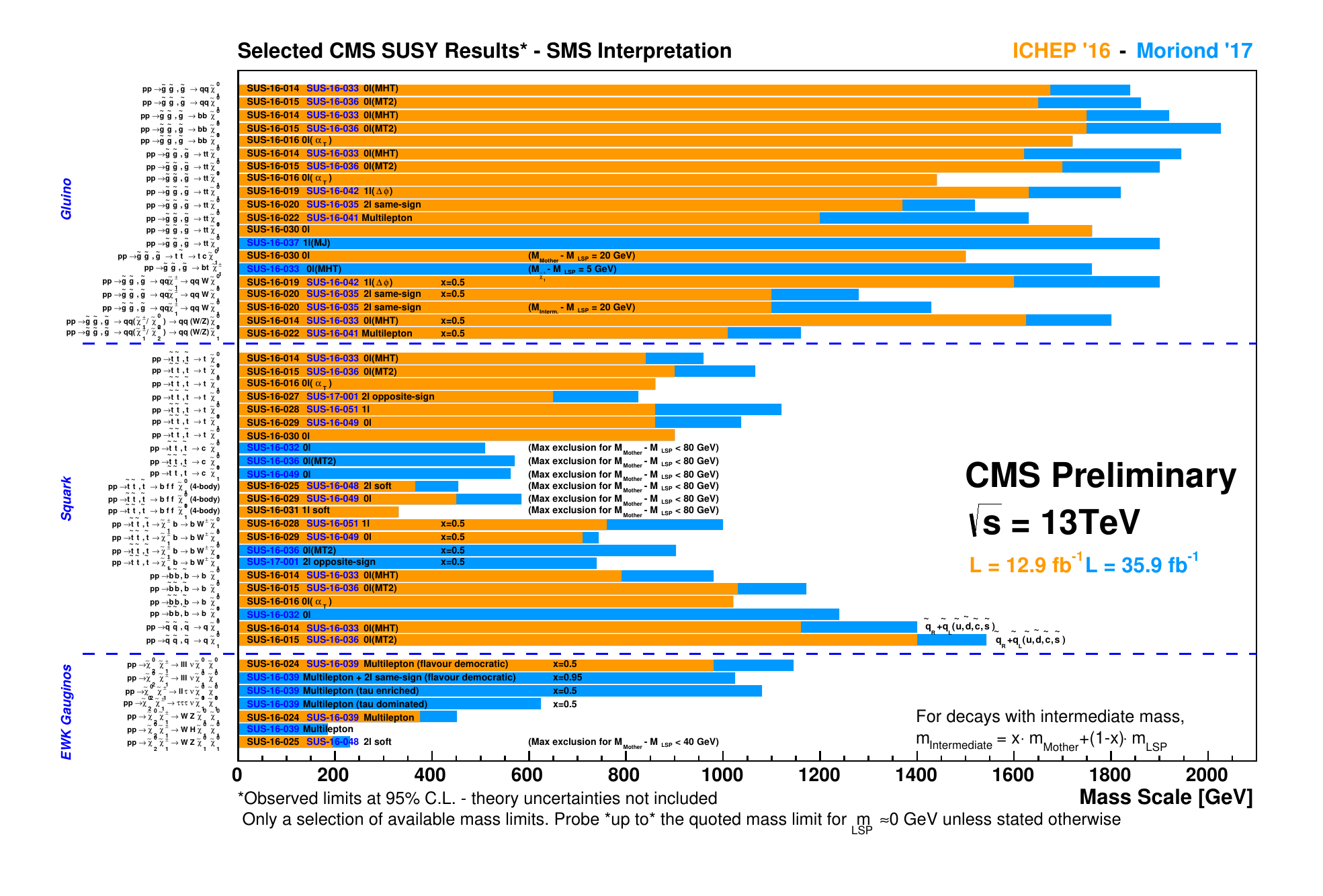}
  \caption{
  Overview of SUSY mass scales probed for various supersymmetric partners. Typical mass scales probed for DM candidates are below $1$ TeV~\cite{cms_summary_susy}.}
  \label{fig:susy_cms}
\end{figure}

So far searches for supersymmetry have not observed any signal while pushing the allowed mass scale very high. Figure~\ref{fig:susy_cms} presents limits on the masses of superpartners. Typical masses for DM candidates are below $1$ TeV, consistent with collider DM searches presented in  Sec.~\ref{sec:collider_dm}.

\begin{figure}[h!]
  \centering 
  \includegraphics[width=.8\textwidth]{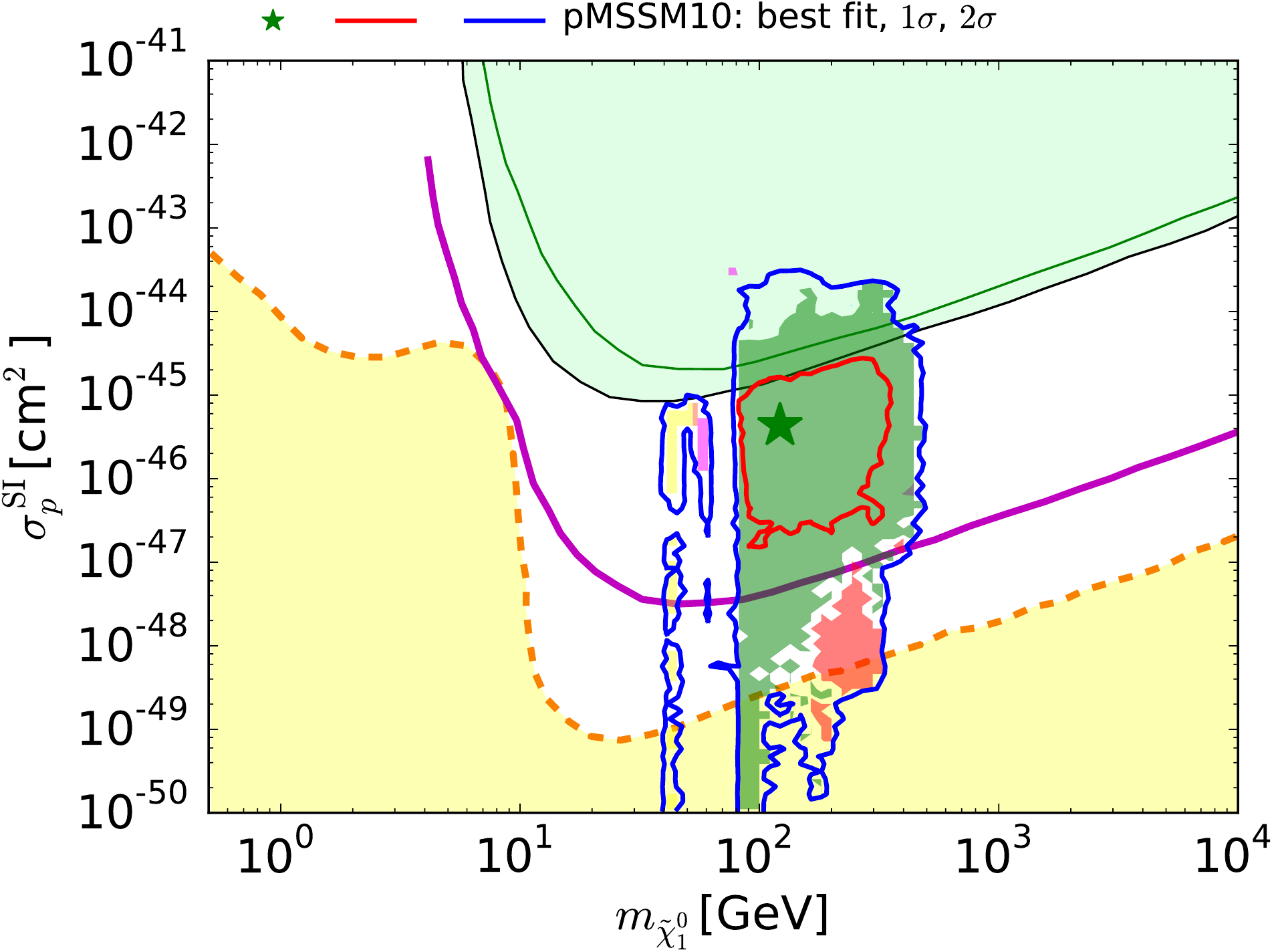}
  \caption{The spin-independent wimp-nucleon plane in the pMSSM10. The red and blue solid lines are the contours for best $\chi^2$ fits, and the solid purple lines show the projected 95\% C.L. exclusion sensitivity of the LZ experiment~\cite{Mount:2017qzi}. The green and black lines show the current sensitivities of the XENON100~\cite{2010PhRvL.105m1302A} and LUX~\cite{Akerib:2016vxi} experiments, respectively, and the dashed orange line shows the astrophysical neutrino ‘floor’~\cite{Cushman:2013zza}, below which astrophysical neutrino backgrounds dominate (yellow region)~\cite{Bagnaschi:2015eha}.
}
\label{fig:susy_pmssm}
\end{figure}

Even at allowed masses, dark matter models need to meet peculiar requirements to still be viable and reproduce the observed relic density in the universe. This may be because of co-annihilation, $s$-channel annihilation or funnel choices.~\cite{Olive:2016efh}.  Phenomenological collaborations performing combined fits to the supersymmetric phase space such as MasterCode and Gambit summarise the most likely phase space of a series of models considering various constraints~\cite{Bagnaschi:2017tru, Costa:2017gup, Athron:2017yua}, see Fig.~\ref{fig:susy_pmssm}.

The dark matter phenomenology in the MSSM is very rich, ranging from  Higgsino-dominated DM annihilation through co-annihilations with other Higgsinos in the early Universe, to resonant annihilation via the light and heavy Higgs funnels, to co-annihilation of neutralinos with both light stops and sbottoms. Combined fits presently prefer  light, Higgsino-dominated neutralinos, with masses of the lightest neutralino to be 
$m_{\tilde{\chi}^0_1}<2.5$ TeV at 95\% CL~\cite{Athron:2017yua}.

\subsection{Higgs to invisible decay}

When considering spin-0 mediated interactions between dark matter and the SM we have to consider the case in which the SM Higgs boson it itself the mediator. Coupling DM to the SM Higgs boson is a natural extension that requires only one additional term to be added to the standard model lagrangian. The Higgs boson might then decay into the DM particles if kinematically accessible, $m_{\rm DM}\leq m_{\rm H}/2 \approx 62.5$~GeV. This allows the production of dark matter in all Higgs production modes. Those are, in descending order of their production cross sections, gluon-fusion (ggH), vector boson-fusion (VBF) associated production with a $W/Z$ boson ($V$H) and associated production with heavy quarks. 

\begin{figure}[h!]
  \centering 
  \includegraphics[width=.6\textwidth]{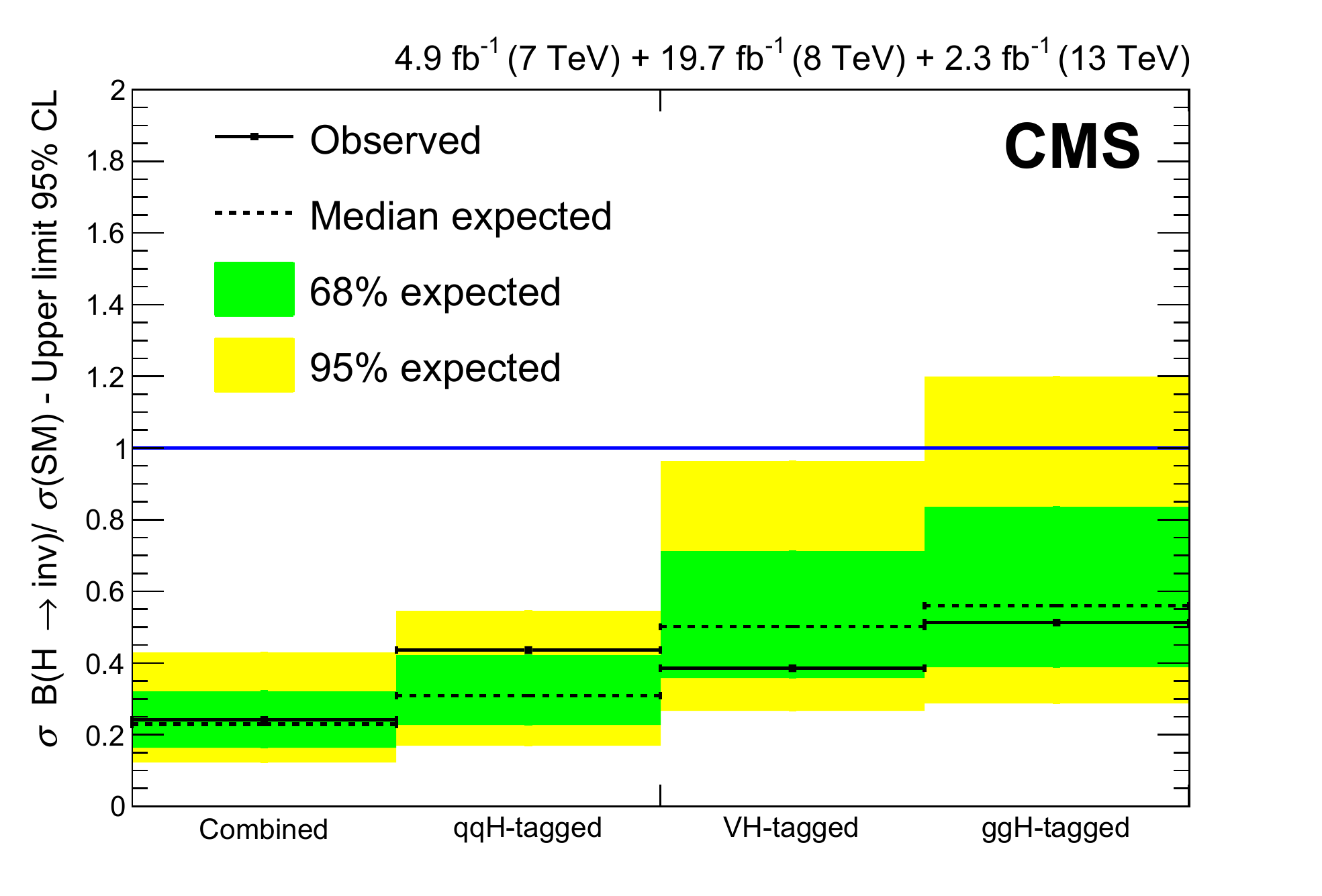}
  \caption{
 Observed and expected 95\% CL limits on $\sigma \times \mathcal{BR}(H \to {\rm inv)}/\sigma(SM)$ for individual combinations of categories targeting gluon-fusion, vector boson-fusion and VH production, and the full combination assuming a Higgs boson with a mass of $m_{\rm H}=125$ GeV~\cite{Khachatryan:2016whc}.}
  \label{fig:h_inv_br}
\end{figure}

Vector boson-fusion is the most sensitive Higgs production mode for $H \to {\rm~inv.}$ decay. The channel is characterised by the presence of two forward jets with a large invariant mass. The cleanest mode is $ZH$ production with the $Z$-boson decaying either in a pair of electrons or muons.  However, this channel has a small cross section times branching fraction, resulting in fewer potential signal events. 
Gluon fusion itself is not directly observable at the LHC because of the overwhelming $QCD$ multijet background. However by requiring that a high momentum ISR jet be produced along with the Higgs boson, one can trigger on the jet and search for the associated invisible Higgs decay, resulting in a signature similar as mono-$j/V$. Several direct measurements of $H \to {\rm inv}$ were also performed~\cite{Khachatryan:2016whc, Aaboud:2017bja, Aad:2015pla}. The combination of the $7, 8$ and $13$~TeV analyses allows upper limits to be set on the Higgs boson to invisible branching fraction $\mathcal{BR}(H \to {\rm inv)}$  observed (expected) 0.24 (0.23) at the 95\% CL. Figure~\ref{fig:h_inv_br} shows the combination and the individual channels. 

\begin{figure}[h!]
  \centering 
  \includegraphics[width=.6\textwidth]{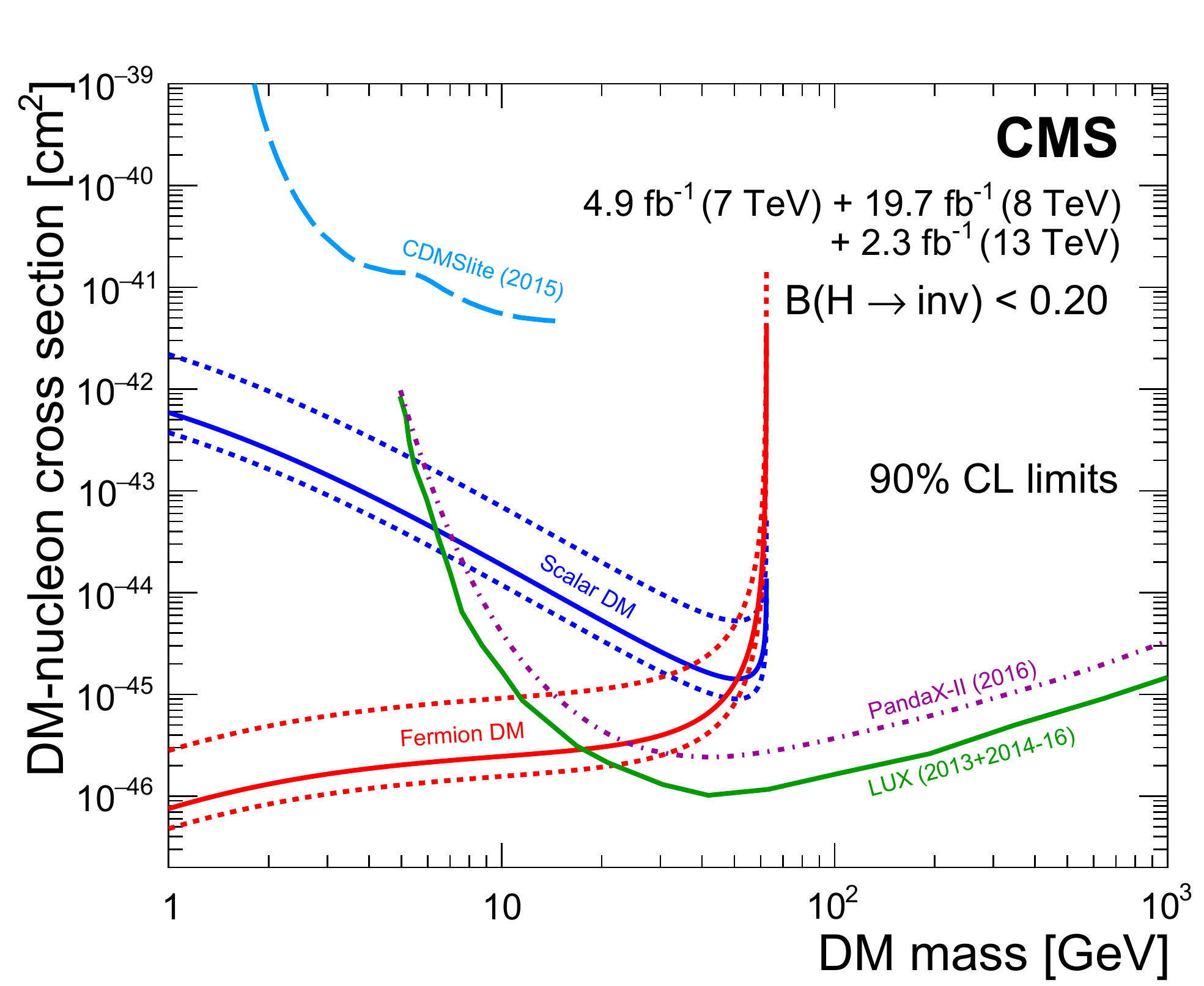}
  \caption{
 Limits on the spin-independent DM-nucleon scattering cross section in Higgs-portal models assuming a scalar or fermion DM particle. The dashed lines show the variation in the exclusion limit using alternative coupling values  as described in~\cite{Khachatryan:2016whc}. The limits are given at the 90\% CL to allow for comparison to direct detection constraints from the LUX~\cite{Akerib:2016vxi}, PandaX-II~\cite{Tan:2016zwf}, and CDMSlite ~\cite{Agnese:2015nto} experiments.}
  \label{fig:h_inv_limits}
\end{figure}

Assuming SM Higgs production cross sections~\cite{Heinemeyer:2013tqa} and tree-level couplings to the SM Higgs sector the upper limit on $\mathcal{BR}(H \to {\rm inv)}$ can be interpreted as an upper limit on the spin-independent DM scattering cross section $\sigma_{\rm SI}$ in direct detection experiments, assuming either  a scalar or fermion DM candidate~\cite{Djouadi:2011aa}. As shown in Fig.~\ref{fig:h_inv_limits} the resulting limits provide stronger constraints for low mass dark matter assuming a scalar fermion DM particle.

\subsection{Low energy and beam dump experiments}
\begin{figure}[h!]
  \centering 
  \includegraphics[width=.6\textwidth]{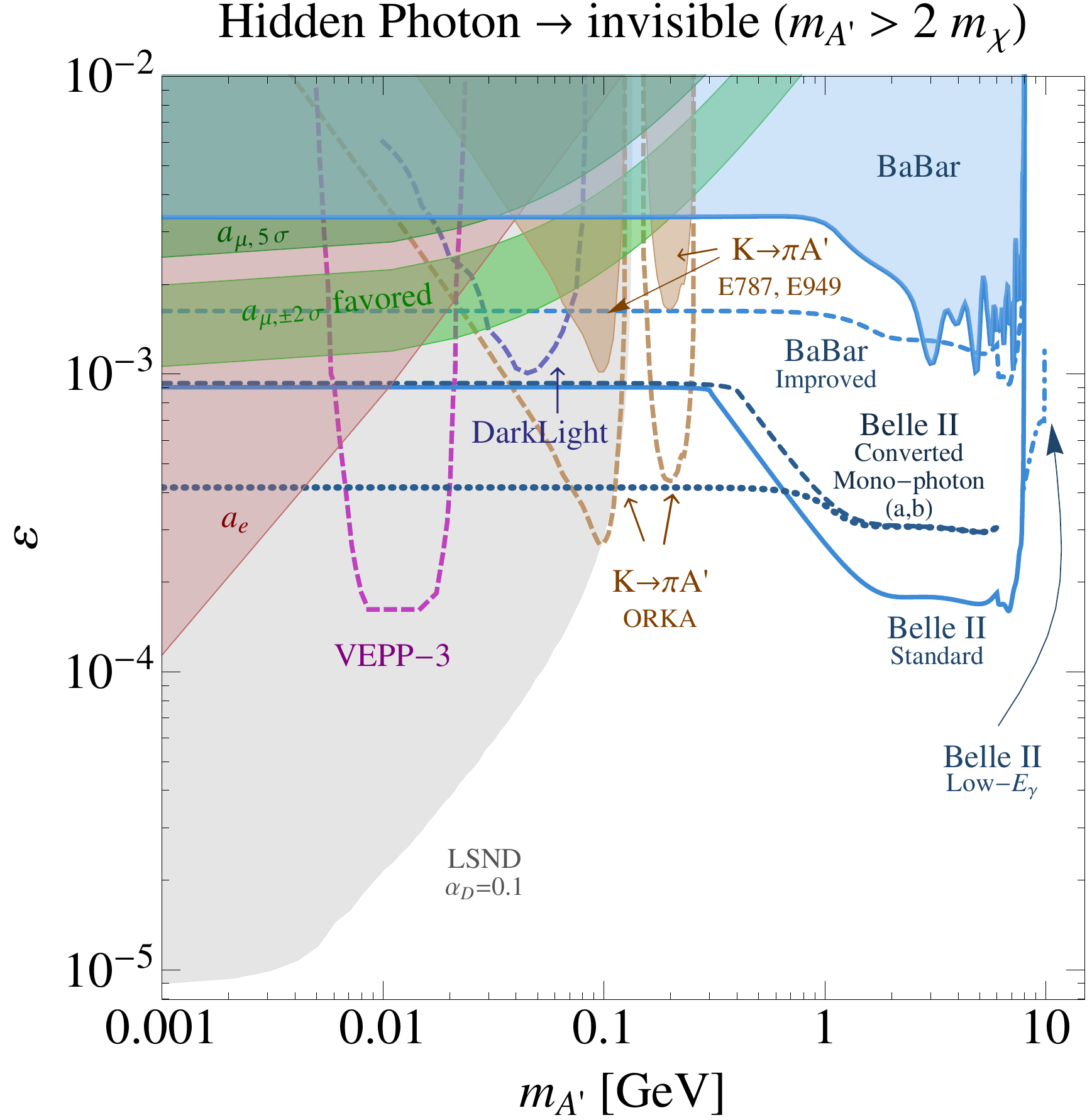}
  \caption{Constrains from the BaBar experiment on light mediators are shown in the light blue shaded region. Possible improvements due to reduced background background, and projections for the Belle-II experiments are shown as blue lines and compared other low energy mediator searches ~\cite{Essig:2013vha}.}
  \label{fig:a_searches}
\end{figure}

Low energy and beam dump experiments have unique sensitivities to well motived low mass mediators. Low energy $e^+ e^-$ data from BaBar has been re-analysed under the light of DM searches. Only about 55~fb$^{-1}$ of the 500~fb$^{-1}$ dataset of BaBar is recorded with a mono-photon trigger. The Belle~\cite{Jaegle:2012sv} and KLOE~\cite{delRio:2016anz} experiments didn't employ a mono-photon trigger and hence no re-analysis is possible. As shown in Fig.~\ref{fig:a_searches} the analysis of BaBar data probes new parameter space and orders of magnitude improvements  will be possible at at Belle-II which is soon to start~\cite{Essig:2013vha}.

\begin{figure}[h!]
  \centering 
  \includegraphics[width=.6\textwidth]{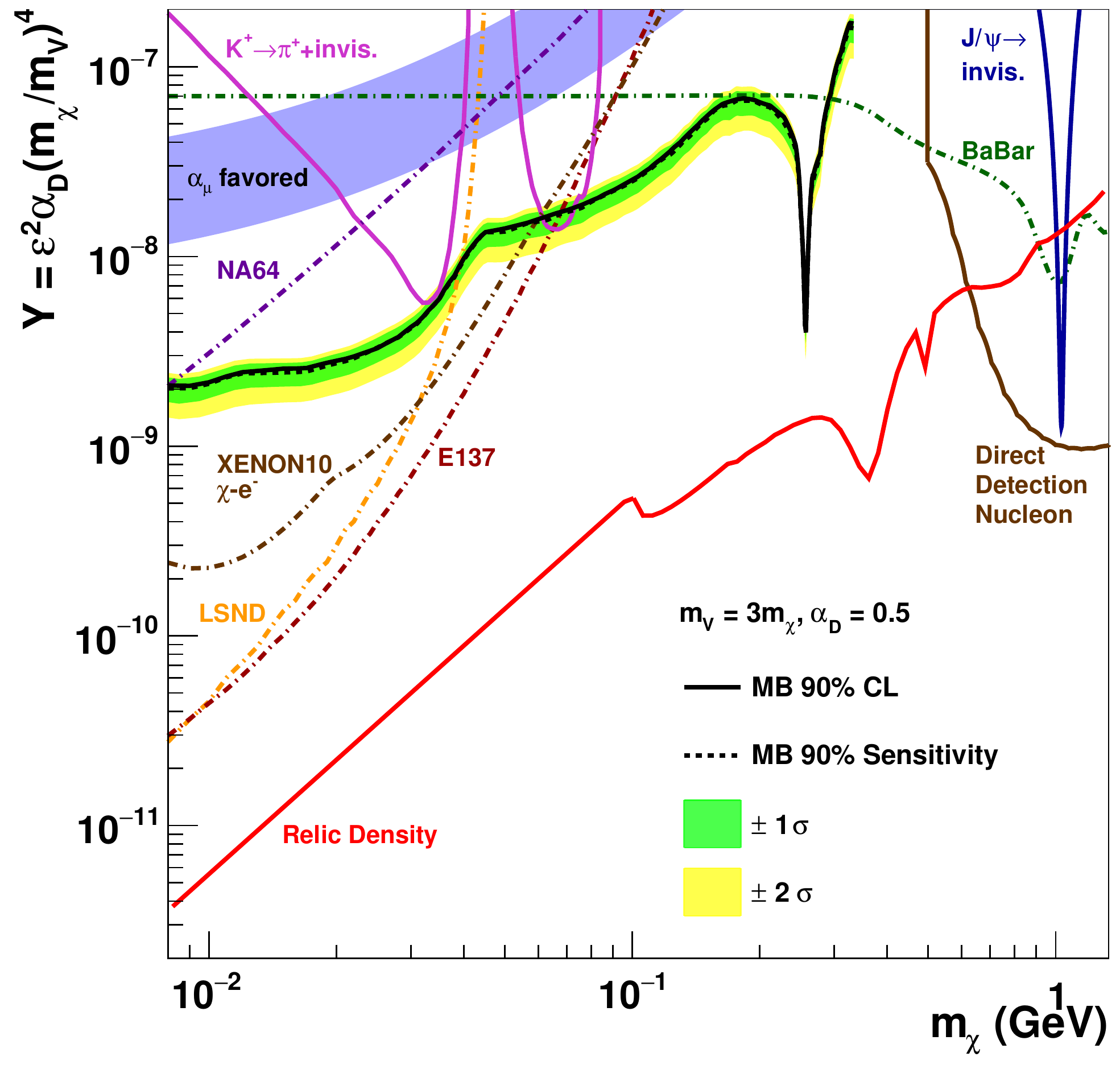}
  \caption{Sensitivity of the Microboone experiment to light mediators compared to other searches. The limits assume DM coupling to quarks/nucleons. The favored parameters to account for the observed relic DM density in this model are shown as the red solid line~\cite{Aguilar-Arevalo:2017mqx}.}
  \label{fig:microboone}
\end{figure}

The MiniBoone~\cite{Aguilar-Arevalo:2017mqx} collaboration is performing a fixed target experiment searching for DM. Employing 8 GeV protons from Fermilab's neutrino beam in 'off-target' mode to reduce neutrino-induced background they search for DM induced event downstream of the target. After accounting for various background the analysis improves upon the BaBar results~\cite{Aguilar-Arevalo:2017mqx}. For vector portal DM model this analysis sets the currently most stringent constraints on low mass DM  of less than $0.3$ GeV, see Fig.~\ref{fig:microboone}.

While proton beam dump experiments are very sensitive to the couplings to quarks, corresponding electron beam dump experiments probe leptonic couplings. Experiments such as APEX~\cite{Abrahamyan:2011gv} and HPS~\cite{Celentano:2014wya} search for dark mediators that are produced by the electron beam and are expected to start data taking in 2018. With several new experiments starting up soon and proposed a more comprehensive coverage of this yet largely unexplored dark matter phase space is expected in the near to medium future~\cite{Battaglieri:2017aum}.

\section{Summary of searches \label{sec:results}}

The vast majority of searches is performed at the LHC on which we focus in this chapter. Low mediator mass searches are scarce and are summarised in Fig.~\ref{fig:microboone}.

\begin{figure}[h!]
  \centering 
  \includegraphics[width=.48\textwidth]{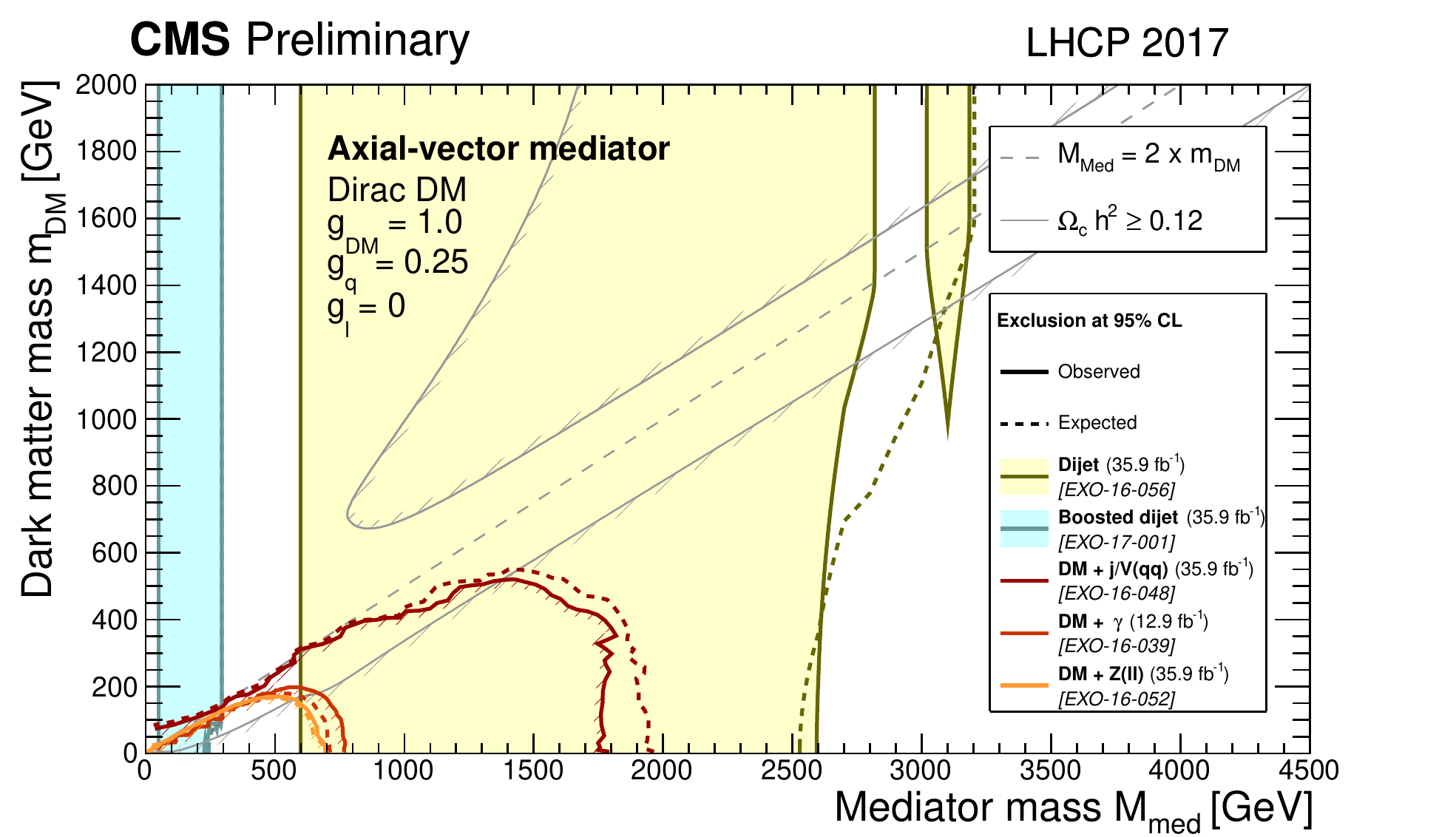}
  \hfill
  \includegraphics[width=.48\textwidth]{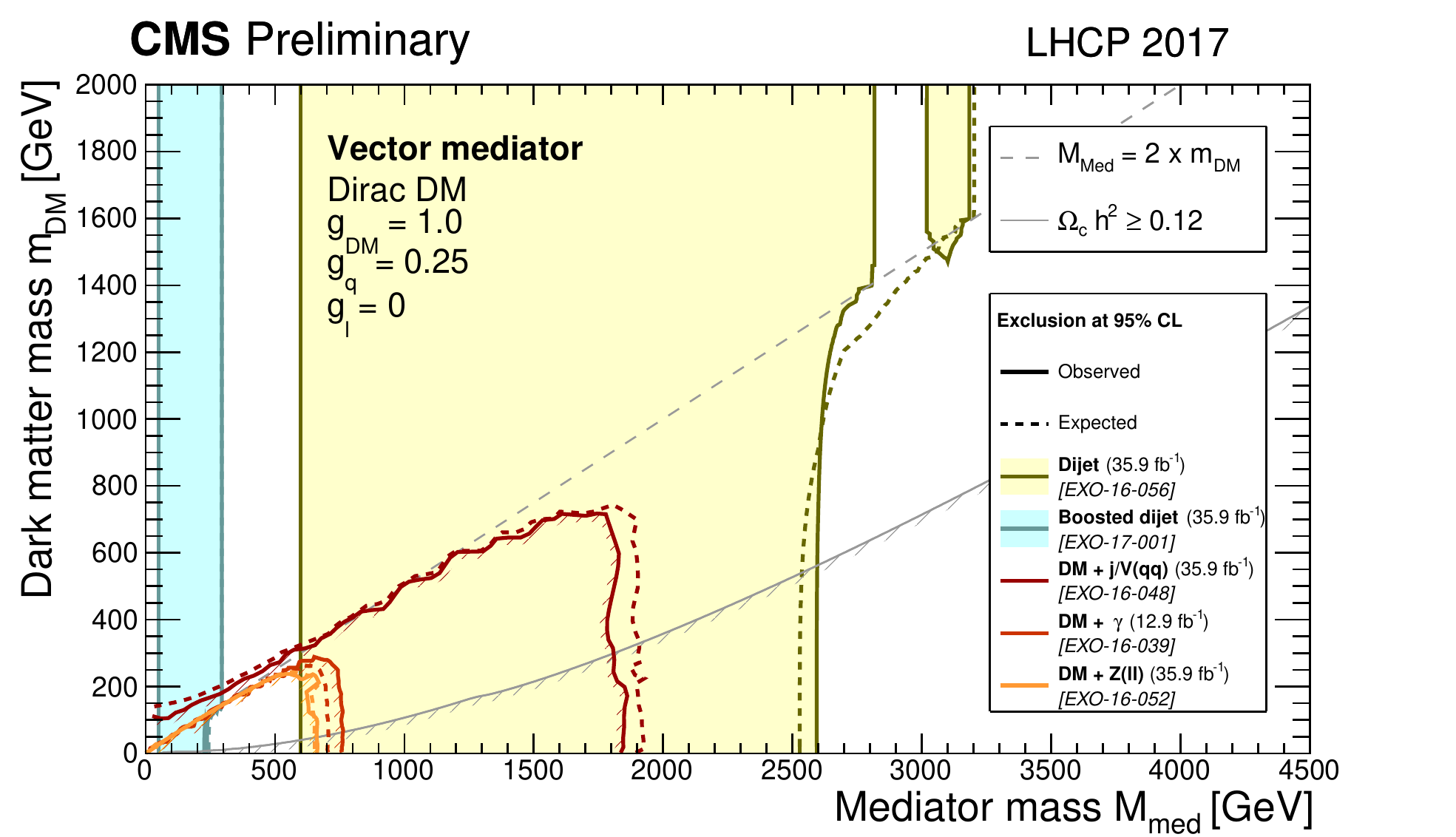}

  \caption{ 95\% CL observed and expected exclusion regions for dijet searches and different $\etmiss$ based DM searches from CMS in the leptophobic axial-vector (left) and the vector (right) model. Following the recommendation of the LHC DM working group, the exclusions are computed for a universal quark coupling $g_{\rm q} = 0.25$ and for a DM coupling of $g_{\rm DM} = 1$. It should be noted that the absolute exclusion of the different searches as well as their relative importance, will strongly depend on the chosen coupling and model scenario. Therefore, the exclusion regions, relic density contours, and unitarity curve shown in this plot are not applicable to other choices of coupling values or model~\cite{cms_summary}.
}
  \label{fig:cms_vector}
\end{figure}

\begin{figure}[h!]
  \centering 
  \includegraphics[width=.48\textwidth]{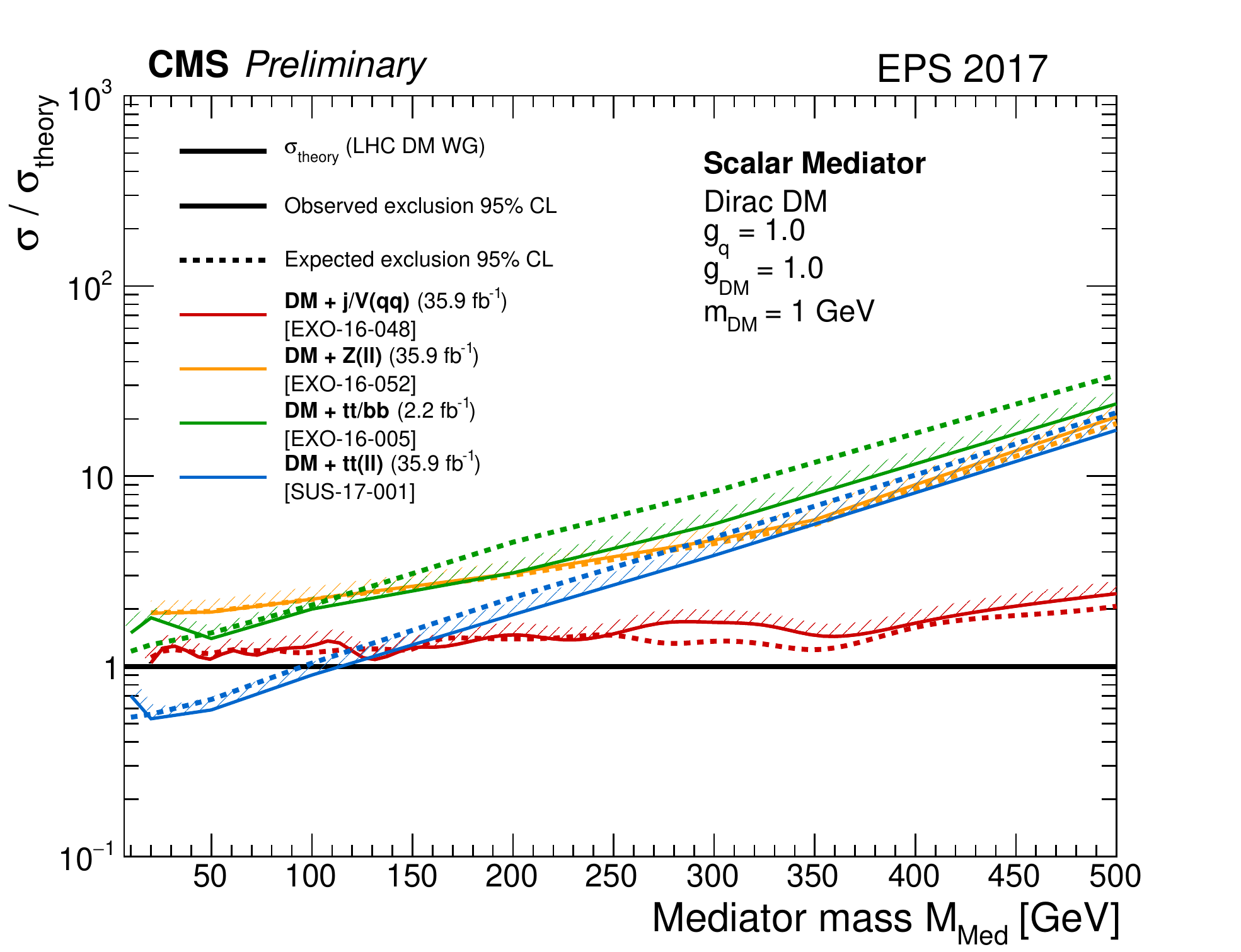}
  \hfill
   \includegraphics[width=.48\textwidth]{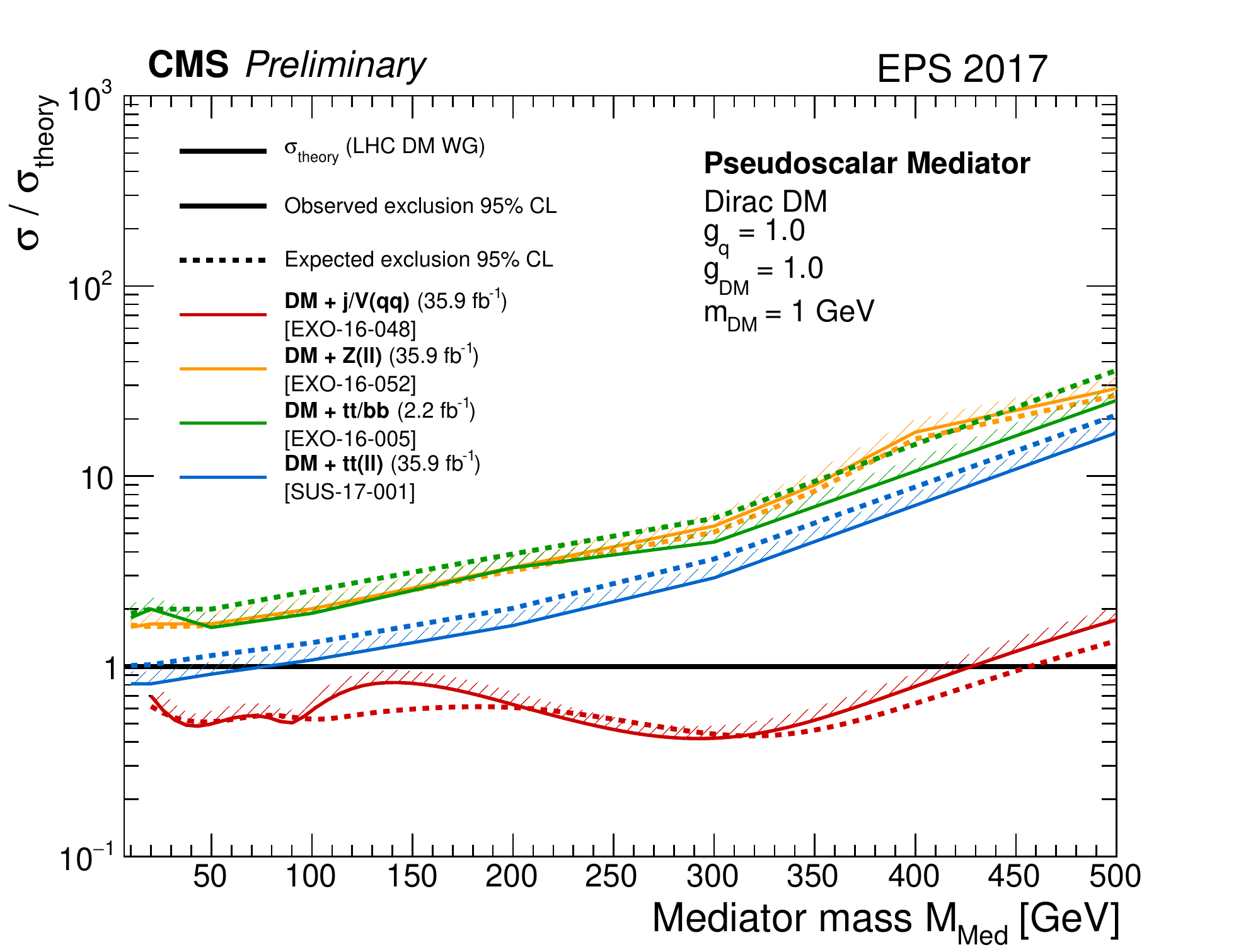}

  \caption{95\% CL observed (solid line) and expected (dashed line) exclusion limits for the scalar
model as a function of $m_{\rm med}$ for different $\etmiss$ based DM searches from CMS. The exclusions are computed for quark coupling $g_{\rm q} = 1.0$ and for a DM coupling of $g_{\rm DM} = 1.0$~\cite{cms_summary}.
}
  \label{fig:cms_scalar}
\end{figure}

\begin{figure}[h!]
  \centering 
  \includegraphics[width=.48\textwidth]{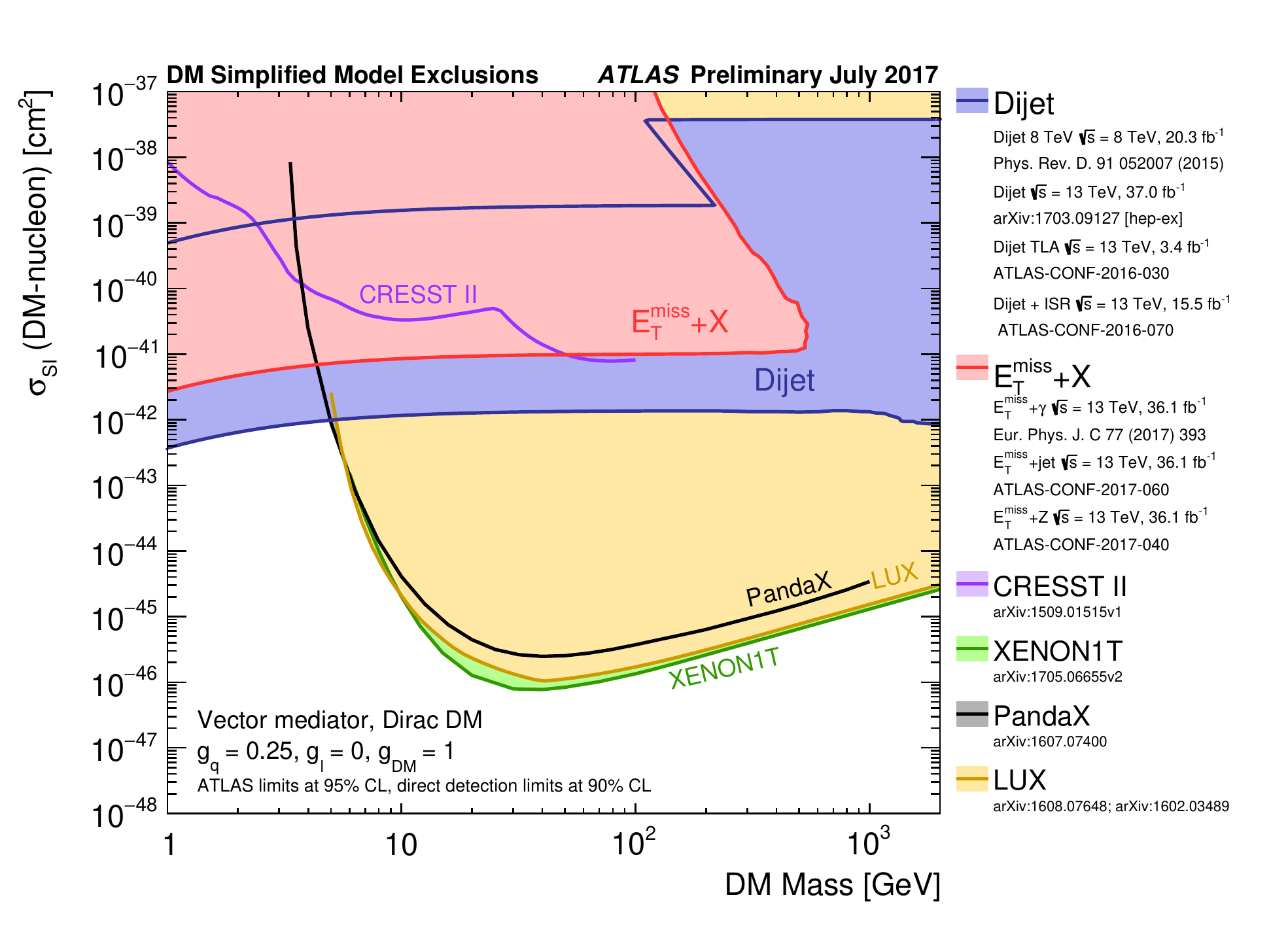}
  \hfill 
\includegraphics[width=.48\textwidth]{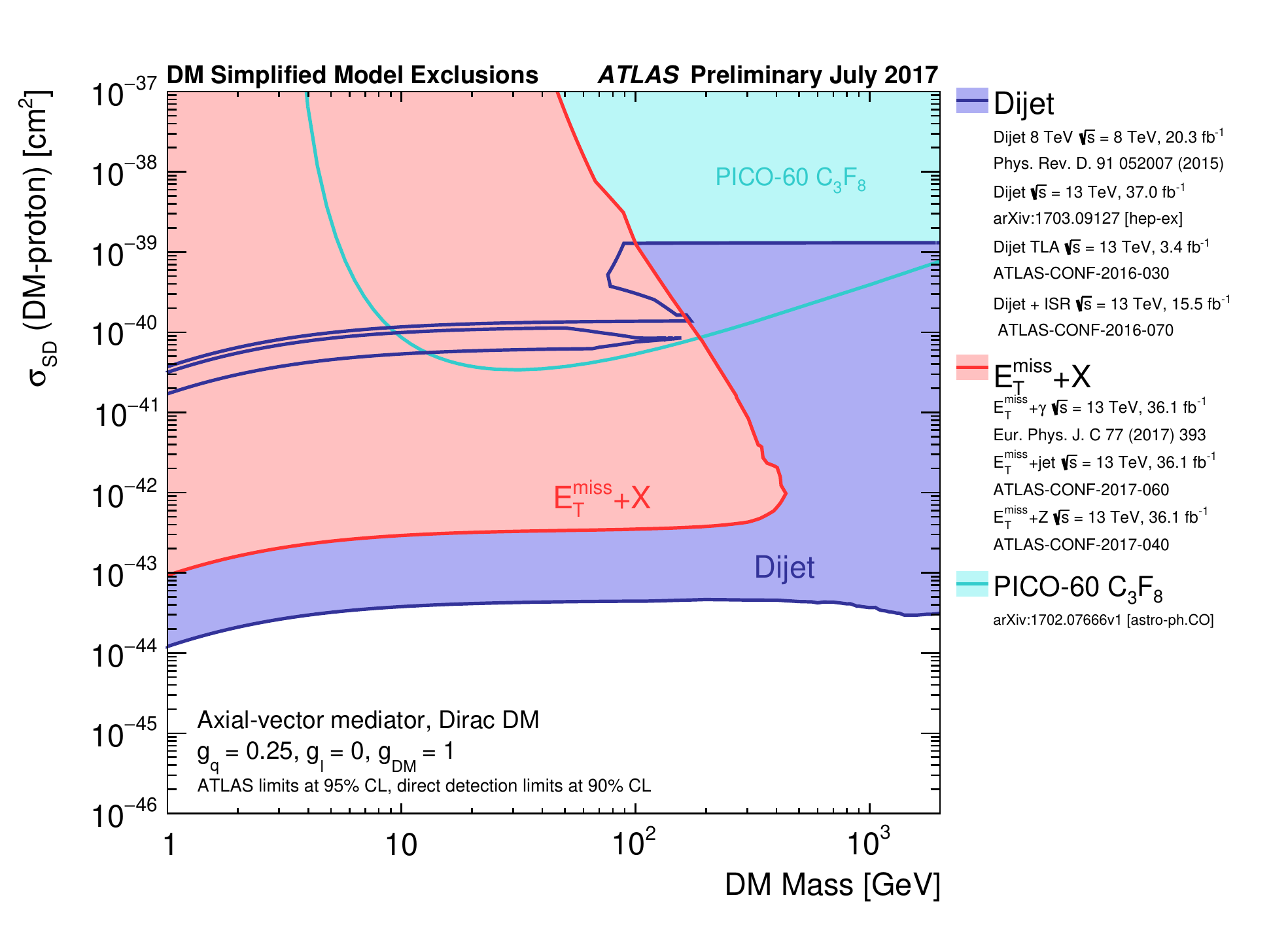}

  \caption{A comparison of the inferred limits to the constraints from direct detection experiments on the spin-independent WIMP-nucleon scattering cross section (left) and spin-dependent WIMP-proton scattering cross section (right) in the context and in the context of the $Z'$-like simplified model with vector (left) and axial-vector couplings. The results from this analysis, excluding the region to the left of the contour, are compared with limits from different direct detection experiments as indicated in the plots. LHC limits are shown at 95\% CL and direct detection limits at 90\% CL. The comparison is valid solely in the context of this model, assuming a mediator width fixed by the dark matter mass and coupling values $g_{\rm q}= 0.25$ and $g_{\rm DM} = 1.0$. LHC searches and direct detection experiments exclude the shaded areas. Exclusions of smaller scattering cross-sections do not imply that larger scattering cross-sections are excluded. The single dijet and $\etmiss+X$ exclusion region represents the union of exclusions from all analyses of that type~\cite{atlas_summary}.
}
  \label{fig:atlas_dd}
\end{figure}

Summary plots of LHC DM searches are taken from the CMS and ATLAS results public  web pages~\cite{cms_summary, atlas_summary}. Despite some differences in analysis strategy and data set sizes, ranging from to $2-36$~fb$^{-1}$,  sensitivities are comparable. In order to do justice to the large effort performed by both collaborations we will present the limits in the $m_{\rm med}-m_{\rm DM}$ mass from the CMS collaboration. The interpretation of the collider results in the spin-independent and spin-dependent WIMP-nucleon planes are taken from the ATLAS collaboration and are shown in Fig.~\ref{fig:atlas_dd}.  The exclusion plots for simplified models using vector and axial-vector couplings are displayed in Fig.~\ref{fig:cms_vector} and Fig.~\ref{fig:cms_scalar} presents the constraints on scalar type couplings.

Present searches given the choice of models and couplings already probe a large parameter space. As displayed in Fig.~\ref{fig:cms_vector} in particular vector and axial-vector exclude up to almost $2$~TeV in mediator mass and correspondingly up to almost 1 TeV in DM mass and a significant fraction of the phase space if applying constraints from the relic density. The  reinterpretation of dijet results extends the exclusion significantly further to about $m_{\rm med}\approx 2.5$~TeV and largely independent of DM mass. The picture is quite different for spin 0 mediator with scalar type interactions, Fig.~\ref{fig:cms_scalar}. Here present searches just start to gain sensitivity in the mass range $m_{\rm med} < 500$ GeV. 

In the context of the spin-independent dark-matter--nucleon-limits used by direct DM experiments and shown in Fig.~\ref{fig:atlas_dd}, left, collider results provide improved sensitivity to low mass DM. While complementary they do not probe new phase space at intermediate to large DM masses.
Collider probe about similar cross sections in the spin-independent $\sigma_{\rm SI}$ and spin-dependent $\sigma_{\rm SD}$ DM mass plane. However, for direct searches the  spin-dependent scattering cross section is not enhanced by coherent scattering over many nucleons and therefore the collider results show improved sensitivities also for larger DM masses as displayed in Fig.~\ref{fig:atlas_dd}, right.

New and updated collider dark matter searches are presented regularly by the CMS and ATLAS collaboration and we want to encourage the reader to look for updates on the experiments' summary pages~\cite{atlas_summary, cms_summary}

\clearpage
\section{Outlook}

Motivated by the WIMP miracle predicting an electroweak origin of dark matter and  providing a rich phenomenology, searching for DM at colliders has become a prominent field.  Collider searches are particularly powerful at low WIMP masses, about the mass of a carbon atom or less, but they also explore larger masses and are not subject to significant astrophysical uncertainties. Direct and indirect searches obtain the best sensitivities at intermediate and large masses but are more dependent on the structure of the interaction (e.g. spin-dependent or independent), astrophysical uncertainties and detector technology. Collider DM searches are therefore a crucial component in the multi-pronged strategy to search for DM and are 
uniquely able to measure various DM properties in case of a discovery. Each of the three  DM detection methods probe different parts of the parameter space with complementary strengths. All three are required to access the full phase space of particle dark matter and eventually measure its properties.

The large experience in analysing the multitude of resulting final states, the rapid increase in centre-of-mass energy and luminosity and a great phenomenological effort had led to an already comprehensive analysis of simple $s$-channel models. 
Despite extending the sensitivity to regions of the phase space that are difficult to access using direct searches, no signal has been found. However, a variety of DM models have been ruled out and new inter-disciplinary developments that will be needed to measure the DM properties instigated.

It is important to note that present models and therefore the corresponding searches are among the most simple ones one can devise. 
For example the known weak currents are a mixture of vector and axial-vector interactions. The successful hunt for the Higgs boson, which is the only fundamental scalar known and theoretically much better understood than DM, also lasted over forty years. Only about 5\% of expected LHC data have been recorded and less than 2\% analysed. Furthermore, the vast majority of searches are exploring a similar $\etmiss+X$ phase space recorded using similar trigger. A discovery might very well still happen, just not as fast as we were hoping given the rapid doubling of luminosity and increase in energy.

In case a discovery takes place first in direct or indirect DM searches, then colliders are best suited to produce and measure DM under laboratory conditions. If the DM is at an energy range not accessible at today's colliders, then DM might constitute a strong case for energy upgrades of the LHC or a future collider~\cite{Mangano:2017tke}.

Dark matter searches are necessarily inter-disciplinary and colliders are one of the corner-stones in our pursuit of DM. As the latest addition to the pantheon of `Exotica' searches they are just starting to mature. New approaches will push the field far beyond today's state. These developments include, but are not restricted to long-lived particle searches~\cite{Banerjee:2017hmw, Buchmueller:2017uqu}, new signatures~\cite{Cohen:2015toa, Banerjee:2017hmw,An:2013xka}, new production modes~\cite{An:2013xka, Bauer:2016gys, Esmaili:2016enf},  dark sectors~\cite{Cohen:2017pzm}, dark-photon searches~\cite{Schmidt-Hoberg:2013hba}, novel detectors~\cite{Curtin:2017izq, Feng:2017uoz, Haas:2014dda} and different beams~\cite{Battaglieri:2014qoa, Abe:2010gxa,Aguilar-Arevalo:2017mqx}. The yet to come 95\% of data, improved  trigger, analysis techniques and more realistic models might very well hold big surprises at the LHC.

\section*{Acknowledgments}
We thank Yangyang Cheng and Ben Krikler for helpful discussions and Carly KleinStern for the careful reading of the manuscript. This work is partially funded and the author iis grateful for the support of the STFC UK, the University of Bristol and Brandeis University.

\clearpage
\bibliographystyle{hunsrt}
\bibliography{review_dm_collider}
\end{document}